\documentclass[preprint]{ptephy_v1}

\preprintnumber{KYUSHU-HET-279, OU-HET-1215} 

\usepackage{graphics}

\usepackage{amsmath} 
\usepackage{amsthm} 
\usepackage{url} 
\usepackage{stmaryrd}
\usepackage{tikz}

\DeclareMathOperator{\tr}{tr}
\DeclareMathOperator{\Tr}{Tr}

\DeclareMathOperator{\interior}{int}
\DeclareMathOperator{\lcm}{lcm}

\newcommand{\Slash}[1]{{\ooalign{\hfil/\hfil\crcr$#1$}}}
\numberwithin{equation}{section}

\allowdisplaybreaks

\begin{document}

\title{Lattice realization of the axial $U(1)$ noninvertible symmetry}


\author[1]{Yamato Honda}
\affil[1]{Department of Physics, Kyushu University, 744 Motooka, Nishi-ku,
Fukuoka 819-0395, Japan}

\author{Okuto Morikawa}
\affil{Department of Physics, Osaka University, Toyonaka, Osaka 560-0043,
Japan}

\author[1]{Soma Onoda}

\author[1]{Hiroshi Suzuki}





\begin{abstract}%
In $U(1)$ lattice gauge theory with compact $U(1)$ variables, we construct the
symmetry operator, i.e.\ the topological defect, for the axial $U(1)$
noninvertible symmetry. This requires a lattice formulation of chiral gauge
theory with an anomalous matter content and we employ the lattice formulation
on the basis of the Ginsparg--Wilson relation. The invariance of the symmetry
operator under the gauge transformation of the gauge field on the defect is
realized, imitating the prescription by Karasik in continuum theory, by
integrating the lattice Chern--Simons term on the defect over \emph{smooth\/}
lattice gauge transformations. The projection operator for allowed magnetic
fluxes on the defect then emerges with lattice regularization. The resulting
symmetry operator is manifestly invariant under lattice gauge transformations.
In an appendix, we give another way of constructing the symmetry operator on
the basis of a 3D $\mathbb{Z}_N$ topological quantum field theory, the
level-$N$ BF theory on the lattice.
\end{abstract}

\subjectindex{B01,B04,B31}

\maketitle

\section{Introduction}
\label{sec:1}
Usually, a symmetry is considered to be lost when it suffers from the anomaly.
Contrary to this common wisdom, it is recently
advocated~\cite{Cordova:2022ieu,Choi:2022jqy} that, in spite of the axial
$U(1)$ anomaly~\cite{Adler:1969gk,Bell:1969ts}, the axial $U(1)$
transformation with fractional rotation angles can be regarded as one form of
the generalized symmetries~\cite{Gaiotto:2014kfa}
(see Refs.~\cite{Schafer-Nameki:2023jdn,Bhardwaj:2023kri,Shao:2023gho} for
reviews), the noninvertible symmetry~\cite{Aasen:2016dop,Bhardwaj:2017xup,Chang:2018iay,Thorngren:2019iar,Komargodski:2020mxz,Koide:2021zxj,Choi:2021kmx,Kaidi:2021xfk,Hayashi:2022fkw,Choi:2022zal,Kaidi:2022uux,Roumpedakis:2022aik,Bhardwaj:2022yxj,Cordova:2022ieu,Choi:2022jqy,Bhardwaj:2022lsg,Karasik:2022kkq,GarciaEtxebarria:2022jky,Choi:2022fgx,Yokokura:2022alv,Nagoya:2023zky,Anber:2023mlc}.
This observation appears to provide a completely new perspective on the notion
of symmetries/anomalies and deserve further investigations.

In the present paper, we make an attempt to implement the above argument of
the axial $U(1)$ noninvertible symmetry on a completely regularized framework
of gauge theory, the lattice gauge theory.\footnote{The present work can be
regarded as a part of our enterprise, the generalized symmetry with lattice
regularization~\cite{Abe:2022nfq,Kan:2023yhz,Abe:2023ncy,Abe:2023ubg,Abe:2023uan,Abe:2023tuf,Abe:2023atv}.} It is well-known that one can construct
a conserved axial $U(1)$ current by adding the Chern--Simons form with an
appropriate coefficient to the axial $U(1)$ current. However, since the
Chern--Simons form is not gauge invariant, the conserved axial $U(1)$ current
cannot be regarded as a physical operator; see Ref.~\cite{Coleman:1978ae} for
this classical issue. In terms of the exponential of the Noether charge defined
on a 3D closed surface~$\mathcal{M}_3$, which is called the symmetry operator
associated with the defect~$\mathcal{M}_3$, $U_\alpha(\mathcal{M}_3)$, where
$\alpha$ is the axial $U(1)$ rotation angle, the addition of the Chern--Simons
form to the axial $U(1)$ current induces the exponential of the Chern--Simons
action on~$\mathcal{M}_3$ with a \emph{noninteger\/} level. Corresponding to
the conservation law of the modified axial $U(1)$ current,
$U_\alpha(\mathcal{M}_3)$ is topological in the sense that it is invariant under
deformations of~$\mathcal{M}_3$. However, since $U_\alpha(\mathcal{M}_3)$ is not
gauge invariant, this way of symmetry restoration had been thought to be
invalid.

The crucial observation of~Refs.~\cite{Cordova:2022ieu,Choi:2022jqy} is that,
when the gauge group is~$U(1)$, the exponential of the Chern--Simons action
with a fractional level can be made gauge invariant by coupling it to a
topological field theory on~$\mathcal{M}_3$. This is analogous to the gauge
invariant low-energy description of the fractional quantum Hall
effect~\cite{Tong:2016kpv}. In this way, the modified symmetry operator,
$\Tilde{U}_\alpha(\mathcal{M}_3)$, which is still topological, can be physical.
However, since $\Tilde{U}_\alpha(\mathcal{M}_3)$ involves a functional integral
for the topological field theory, $\Tilde{U}_\alpha(\mathcal{M}_3)$ is not
unitary; it is rather an infinite sum of unitary operators. Thus, the
resulting symmetry is noninvertible because the Hermitian conjugate
of~$\Tilde{U}_\alpha(\mathcal{M}_3)$ does not provide the inverse.

To realize the above axial $U(1)$ noninvertible symmetry on the lattice,
therefore, we have to do mainly two things: One is to formulate the lattice
fermion which couples to the defect defined on the lattice through the
$\gamma_5$ coupling. This is because the deformation of the defect induces the
axial $U(1)$ transformation on the fermion; this is nothing but the axial
$U(1)$ Ward--Takahashi identity. The other thing is to formulate the $U(1)$
Chern--Simons action on the defect; we also have to implement the gauge
invariance of the Chern--Simons action in a certain way. These two things are
of course mutually related through the axial $U(1)$ anomaly and both tasks are
not straightforward.

For the first task, it is natural to regard the defect as a particular
configuration of an external $U(1)$ gauge field (for the details, see below).
The lattice fermion couples to this $U(1)$ lattice gauge field through the
$\gamma_5$ coupling. This thus requires a formulation of \emph{chiral gauge
theory\/} on the lattice, which itself is known to be a very hard problem.
Moreover, the matter content of our chiral gauge theory is \emph{anomalous},
because the gauge anomaly in the chiral gauge theory corresponds to the axial
$U(1)$ anomaly in the target vector-like theory. For this, we employ the
lattice formulation of the $U(1)$ chiral gauge theory~\cite{Luscher:1998du} on
the basis of the Ginsparg--Wilson relation~\cite{Ginsparg:1981bj,Neuberger:1997fp,Hasenfratz:1997ft,Hasenfratz:1998ri,Neuberger:1998wv,Hasenfratz:1998jp,Luscher:1998pqa,Niedermayer:1998bi}.
Remarkably, in this lattice formulation, one can determine the structure of the
axial $U(1)$ anomaly with finite lattice
spacings~\cite{Luscher:1998kn,Fujiwara:1999fi}.

The second task, the Chern--Simons term on the lattice is also a quite
nontrivial problem. One possible approach is to use the Villain-type
noncompact $U(1)$ variable as in a recent work on the $U(1)$ Chern--Simons
action on the lattice~\cite{Jacobson:2023cmr}. In the present paper, we
consider a formulation on the basis of the compact $U(1)$ variables with a
possible generalization to non-Abelian gauge theories in mind. Then, it appears
difficult to imitate the procedure
in~Refs.~\cite{Cordova:2022ieu,Choi:2022jqy} directly, because it relies on a
change of integration variables in terms of the gauge potential. Instead, we
find that a formulation of~Ref.~\cite{Karasik:2022kkq}, which is based on a
``gauge average,'' is rather suitable for our lattice formulation. See also
Ref.~\cite{GarciaEtxebarria:2022jky} for an analogous construction. Imitating
the prescription of~Ref.~\cite{Karasik:2022kkq}, we thus integrate the
Chern--Simons term on the defect, which is defined so as to cancel the axial
$U(1)$ anomaly in our lattice formulation, over \emph{smooth\/} lattice gauge
transformations on the defect. Then, as per~Ref.~\cite{Karasik:2022kkq}, the
projection operator for allowed magnetic fluxes on the defect emerges. The
resulting expression is manifestly invariant under lattice gauge
transformations. In this way, we realize the axial $U(1)$ noninvertible
symmetry with lattice regularization.

This paper is organized as follows: In Sect.~\ref{sec:2}, we give a
general setup of our lattice formulation. We introduce two lattice gauge
fields, one is for the physical and dynamical $U(1)$ gauge field and the other
is for the external $U(1)$ gauge field which defines the defect.
Section~\ref{sec:3} is a rather lengthy exposition of our construction of the
lattice fermion with desired properties. We admit that our lattice formulation
appears ``too heavy'' just in order to describe a simple system of a massless
Dirac fermion coupled to the dynamical $U(1)$ gauge field. However, since the
lattice fermion possesses the $\gamma_5$~coupling to the (albeit external)
gauge field, we think that this formulation cannot be readily simplified. The
most important result in~Sect.~\ref{sec:3} is~Eq.~\eqref{eq:(3.31)}, the axial
$U(1)$ Ward--Takahashi identity on the lattice, and, for the construction of
the symmetry operator in~Sect.~\ref{sec:4}, it is sufficient to accept this
relation; the purpose of the rest of~Sect.~\ref{sec:3} is to show the existence
of a fermion integration measure which ensures~Eq.~\eqref{eq:(3.31)}.
In~Sect.~\ref{sec:4.1}, on the basis of the formulation in~Sect.~\ref{sec:3},
we define the topological defect as a particular configuration of the external
$U(1)$ lattice gauge field. The Chern--Simons term on the defect emerges as the
surface term of the volume integral of the axial $U(1)$ anomaly. Then,
in~Sect.~\ref{sec:4.2}, imitating the prescription
of~Ref.~\cite{Karasik:2022kkq}, we integrate the Chern--Simons term on the
defect over gauge transformations on the defect. Here, to reproduce the picture
assumed in~Ref.~\cite{Karasik:2022kkq}, it is important to carry the
integration only over \emph{smooth\/} lattice gauge
transformations.\footnote{Though we have to assume an elaborated form of the
integration measure; see Appendix~\ref{sec:B}.} Then, the projection operator
for allowed magnetic fluxes on the defect emerges. We note that the resulting
symmetry operator as the whole is manifestly invariant under lattice gauge
transformations. Section~\ref{sec:5} is devoted to Conclusion.
In~Appendix~\ref{sec:A}, we elucidate the issue of the local counterterm which
provides a desired form of the gauge anomaly; the information obtained in this
appendix is utilized in the construction of our lattice fermion
in~Sect.~\ref{sec:3}. In~Appendix~\ref{sec:B}, we provide the precise
definition of the integration measure for the smooth gauge degrees of freedom.
In~Appendix~\ref{sec:C},\footnote{We are deeply indebted to Yuya Tanizaki for
the contents of~Appendix~\ref{sec:C}.} we give another way of constructing the
symmetry operator for the rotation angle~$\alpha=2\pi p/N$ with even~$p$, on
the basis of a 3D $\mathbb{Z}_N$ topological quantum field theory (TQFT), the
level-$N$ BF theory on the lattice. This construction resolves some
unsatisfactory points in the construction in the main text. The construction is
intrinsically 3D and can also be applied to homologically nontrivial closed
3-surfaces; the symmetry operator is manifestly invariant under 3D lattice
gauge transformations. The computation of fusion rules of symmetry operators is
also given.

\section{$U(1)\times U(1)'$ lattice gauge theory}
\label{sec:2}
We consider a lattice gauge theory defined on a finite hypercubic lattice of
size~$L$:\footnote{The Lorentz index is denoted by Greek letters $\mu$, $\nu$,
\dots, and runs over 0, 1, 2, and~3.}
\begin{equation}
   \Gamma
   =\left\{x\in\mathbb{Z}^4\mathrel{}\middle|\mathrel{}0\leq x_\mu<L\right\}.
\label{eq:(2.1)}
\end{equation}
The gauge group is~$U(1)\times U(1)'$.\footnote{Reference~\cite{Kadoh:2007xb}
was quite useful in considering a lattice formulation of the Weyl fermion with
a product gauge group.} The first $U(1)$ factor is for a physical and dynamical
gauge field and the latter $U(1)'$ is for an external nondynamical gauge
field. Corresponding to these, we introduce two $U(1)$ lattice gauge fields by
link variables,\footnote{In this paper, we basically denote quantities
associated with dynamical fields, which are subject of the functional
integrals, by lowercase letters, while quantities associated with external
nondynamical fields by capital letters.}
\begin{equation}
   u(x,\mu)\in U(1),\qquad
   U(x,\mu)\in U(1),
\label{eq:(2.2)}
\end{equation}
where both obey periodic boundary conditions:\footnote{$\Hat{\mu}$ denotes the
unit vector in direction~$\mu$.}
\begin{equation}
   u(x+L\Hat{\nu},\mu)=u(x,\mu),\qquad
   U(x+L\Hat{\nu},\mu)=U(x,\mu).
\label{eq:(2.3)}
\end{equation}
$u(x,\mu)$ is the dynamical $U(1)$ gauge field which couples to fermion fields
in a vector-like way. The external $U(1)$ gauge field $U(x,\mu)$, on the other
hand, couples to fermion fields in a chirally asymmetric, i.e.\ through
$\gamma_5$, way. See the next section. The sole role of the external gauge
field~$U(x,\mu)$ is to define the topological defect associated with the axial
$U(1)$ (noninvertible) symmetry. In this paper, we consider only a
homologically trivial 3D defect~$\widetilde{M}_3$ on the dual lattice
of~$\Gamma$. As we will see below, $U(x,\mu)$ is given by a lattice analogue of
the Poincar\'e dual of~$\widetilde{M}_3$ and we can assume that $U(x,\mu)$ is
pure-gauge:
\begin{equation}
   U(x,\mu)=\Lambda(x)^{-1}\Lambda(x+\Hat{\mu}),\qquad\Lambda(x)\in U(1),
\label{eq:(2.4)}
\end{equation}
where $\Lambda(x+L\hat{\mu})=\Lambda(x)$.

Fermion fields, whose detailed description is deferred to the next section, are
supposed to be residing on sites of~$\Gamma$. The expectation value is then
defined by the functional integral,
\begin{equation}
   \langle\mathcal{O}\rangle
   =\frac{1}{\mathcal{Z}}
   \int\mathrm{D}[u]\,e^{-S_\mathrm{G}}\,
   \left\langle\mathcal{O}\right\rangle_{\mathrm{F}},\qquad
   \langle\mathcal{O}\rangle_{\mathrm{F}}
   =w[m]\int\mathrm{D}[\psi]\mathrm{D}[\Bar{\psi}]\,
   e^{-S_\mathrm{F}}\,\mathcal{O},
\label{eq:(2.5)}
\end{equation}
where the partition function $\mathcal{Z}$ is determined by requiring
$\langle1\rangle=1$.

The integral over the dynamical $U(1)$ gauge field~$u(x,\mu)$ is defined by the
Haar measure at each link as usual. We adopt, however, a somewhat
unconventional gauge action~$S_{\mathrm{G}}$
following~Ref.~\cite{Luscher:1998du},
\begin{equation}
   S_{\mathrm{G}}=\frac{1}{4g_0^2}\sum_{x\in\Gamma}\sum_{\mu,\nu}
   \mathcal{L}_{\mu\nu}(x)
\label{eq:(2.6)}
\end{equation}
with $g_0$ being the bare coupling and, for a fixed number~$0<\epsilon<\pi/3$,
\begin{equation}
   \mathcal{L}_{\mu\nu}(x)
   =\begin{cases}
   [f_{\mu\nu}(x)]^2\left\{1-[f_{\mu\nu}(x)]^2/\epsilon^2\right\}^{-1}&
   \text{if $|f_{\mu\nu}(x)|<\epsilon$},\\
   \infty&\text{otherwise}.\\
   \end{cases}
\label{eq:(2.7)}
\end{equation}
In this expression, the field strength of~$u(x,\mu)$ is defined by
\begin{equation}
   f_{\mu\nu}(x)
   =\frac{1}{i}\ln p(x,\mu,\nu),\qquad
   -\pi<f_{\mu\nu}(x)\leq\pi,
\label{eq:(2.8)}
\end{equation}
from the plaquette variable,
\begin{equation}
   p(x,\mu,\nu)
   =u(x,\mu)u(x+\Hat{\mu},\nu)u(x+\Hat{\nu},\mu)^{-1}u(x,\nu)^{-1}.
\label{eq:(2.9)}
\end{equation}
The gauge action~\eqref{eq:(2.6)} is designed to impose the restriction, the
so-called admissibility, on possible gauge field configurations:
\begin{equation}
   \sup_{x,\mu,\nu}|f_{\mu\nu}(x)|<\epsilon.
\label{eq:(2.10)}
\end{equation}
This is a gauge-invariant smoothness condition on the gauge field.\footnote{If
we restore the lattice spacing~$a$, the admissibility reads
$|f_{\mu\nu}(x)|<\epsilon/a^2$ and, in the continuum limit~$a\to0$, this does
not place any restriction on differentiable gauge fields.} In lattice gauge
theory (with compact variables), the admissibility plays crucial roles to
define topological quantities in a transparent
way~\cite{Luscher:1981zq,Hernandez:1998et}. For instance, in the present $U(1)$
lattice gauge theory, the space of admissible gauge fields is given by a
disjoint union of topological sectors, each of which is labeled by the
magnetic fluxes~\cite{Luscher:1998du},
\begin{equation}
   m_{\mu\nu}
   =\frac{1}{2\pi}
   \sum_{s,t=0}^{L-1}f_{\mu\nu}(x+s\Hat{\mu}+t\Hat{\nu}).
\label{eq:(2.11)}
\end{equation}
The fermion integration measure in~Eq.~\eqref{eq:(2.5)} is thus defined for
each magnetic flux sector separately. The factor~$w[m]$ in~Eq.~\eqref{eq:(2.5)}
parametrizes the relative weight and phase for each magnetic flux sector
specified by~$m_{\mu\nu}$.

We note that if we parametrize the $U(1)$ gauge field as
\begin{equation}
   u(x,\mu)=e^{ia_\mu(x)},\qquad
   -\pi<a_\mu(x)\leq\pi,
\label{eq:(2.12)}
\end{equation}
then the field strength~\eqref{eq:(2.8)} is written
as\footnote{$\partial_\mu f(x)=f(x+\Hat{\mu})-f(x)$ is the forward difference
operator.}
\begin{equation}
   f_{\mu\nu}(x)
   =\partial_\mu a_\nu(x)-\partial_\nu a_\mu(x)+2\pi z_{\mu\nu}(x),
\label{eq:(2.13)}
\end{equation}
where $z_{\mu\nu}(x)\in\mathbb{Z}$. Note that $z_{\mu\nu}(x)$ is generally
non-zero because of a possible mismatch between the logarithmic branch
in~Eq.~\eqref{eq:(2.8)} and that for the product of~Eq.~\eqref{eq:(2.12)}. From
this, we have\footnote{Unless stated otherwise, it is understood that repeated
indices are summed over.}
\begin{equation}
   \left|
   \frac{1}{2}\varepsilon_{\mu\nu\rho\sigma}
   \partial_\nu z_{\rho\sigma}(x)
   \right|
   =\left|
   \frac{1}{4\pi}\varepsilon_{\mu\nu\rho\sigma}
   \partial_\nu f_{\rho\sigma}(x)
   \right|
   <\frac{12}{4\pi}\epsilon<1,
\label{eq:(2.14)}
\end{equation}
owing to~Eqs.~\eqref{eq:(2.13)} and~\eqref{eq:(2.10)}. Since the most left-hand
side of this inequality is a sum of (six) integers, it must identically vanish,
\begin{equation}
   \varepsilon_{\mu\nu\rho\sigma}
   \partial_\nu z_{\rho\sigma}(x)=0,
\label{eq:(2.15)}
\end{equation}
i.e.\ the integer field~$z_{\rho\sigma}(x)$ in~Eq.~\eqref{eq:(2.13)} satisfies
the Bianchi identity. This also implies that the field
strength~$f_{\mu\nu}(x)$~\eqref{eq:(2.13)} itself satisfies the Bianchi
identity,
\begin{equation}
   \varepsilon_{\mu\nu\rho\sigma}
   \partial_\nu f_{\rho\sigma}(x)=0.
\label{eq:(2.16)}
\end{equation}

For the the external gauge field $U(x,\mu)$, on the other hand, it is
pure-gauge~\eqref{eq:(2.4)} and the corresponding field strength identically
vanishes:
\begin{equation}
   F_{\mu\nu}(x)
   =\frac{1}{i}\ln
   \left[U(x,\mu)U(x+\Hat{\mu},\nu)U(x+\Hat{\nu},\mu)^{-1}U(x,\nu)^{-1}\right]
   =0.
\label{eq:(2.17)}
\end{equation}
We will use this fact frequently in what follows.

\section{Fermion sector}
\label{sec:3}
The definition of the fermion integration in~Eq.~\eqref{eq:(2.5)} requires an
elaborate consideration as done in~Ref.~\cite{Luscher:1998du}. This is
because the fermion fields couples to the external gauge field $U(x,\mu)$ in a
chirally asymmetric way and this requires a construction of chiral gauge theory
on the lattice, a hard problem. Moreover, the $U(1)'$ sector possesses the
gauge anomaly to reproduce the axial $U(1)$ anomaly in the target vector-like
theory. Therefore, we have to define an ``anomalous gauge theory'' on
the lattice.\footnote{This gauge anomaly is physically harmless, because
$U(x,\mu)$ is not a dynamical gauge field.} Having the results
of~Ref.~\cite{Luscher:1998du}, however, our task to define the fermion
integration measure is not so quite hard, because the gauge field which causes
the chiral coupling, $U(x,\mu)$, is not dynamical and moreover pure-gauge,
Eq.~\eqref{eq:(2.4)}.

\subsection{General setting and the lattice Dirac operator}
\label{sec:3.1}
We consider a multiplet of two left-handed \emph{Weyl\/} fermion fields,
$\psi_\alpha$ ($\alpha=1$, 2 is the ``flavor'' index), each possesses $U(1)$
charges under the gauge group~$U(1)\times U(1)'$:
\begin{equation}
   (\mathrm{e}_1,\mathrm{e}_1')=(+1,-1),\qquad
   (\mathrm{e}_2,\mathrm{e}_2')=(-1,-1).
\label{eq:(3.1)}
\end{equation}
Since $\psi$ is Weyl, the coupling to~$U(1)'$ is chiral, whereas the coupling
to~$U(1)$ is vector-like (as per the Dirac fermion in quantum electrodynamics).
The forward and backward covariant difference operators on fermion fields are
thus defined by
\begin{align}
   \nabla_\mu\psi(x)
   &=R[u(x,\mu)]R[U(x,\mu)]\psi(x+\Hat{\mu})-\psi(x),
\notag\\
   \nabla_\mu^*\psi(x)
   &=\psi(x)
   -R[u(x-\Hat{\mu},\mu)]^{-1}R[U(x-\Hat{\mu},\mu)]^{-1}\psi_\alpha(x-\Hat{\mu}),
\label{eq:(3.2)}
\end{align}
where
\begin{equation}
   R[u(x,\mu)]_{\alpha\beta}
   =\delta_{\alpha\beta}u(x,\mu)^{\mathrm{e}_\alpha},\qquad
   R[U(x,\mu)]_{\alpha\beta}
   =\delta_{\alpha\beta}U(x,\mu)^{\mathrm{e}_\alpha'}.
\label{eq:(3.3)}
\end{equation}
From these, the Wilson Dirac operator is given by
\begin{equation}
   D_{\mathrm{w}}=\frac{1}{2}
   \left[\gamma_\mu(\nabla_\mu+\nabla_\mu^*)-\nabla_\mu\nabla_\mu^*\right].
\label{eq:(3.4)}
\end{equation}
In our construction, we assume a lattice Dirac operator which fulfills the
Ginsparg--Wilson relation~\cite{Ginsparg:1981bj},
\begin{equation}
   \gamma_5D+D\gamma_5=D\gamma_5D
\label{eq:(3.5)}
\end{equation}
with the chiral matrix~$\gamma_5$. The overlap Dirac
operator~\cite{Neuberger:1997fp,Neuberger:1998wv} given by
\begin{equation}
   D=1-A(A^\dagger A)^{-1/2},\qquad A=1-D_{\mathrm{w}}
\label{eq:(3.6)}
\end{equation}
is an explicit example of such a lattice Dirac operator. From the above
expressions, one sees the $\gamma_5$-hermiticity, $D^\dagger=\gamma_5D\gamma_5$.

The gauge covariance of the lattice difference operators~\eqref{eq:(3.2)} is
inherited by~$D$. Under the $U(1)\times U(1)'$ gauge transformation,
\begin{equation}
   u(x,\mu)\to\lambda(x)u(x,\mu)\lambda(x+\Hat{\mu})^{-1},\qquad
   U(x,\mu)\to\Lambda(x)U(x,\mu)\Lambda(x+\Hat{\mu})^{-1},
\label{eq:(3.7)}
\end{equation}
where $\lambda(x)\in U(1)$ and~$\Lambda(x)\in U(1)$, then we have
\begin{equation}
   D\to
   R[\lambda]R[\Lambda]
   DR[\lambda]^{-1}R[\Lambda]^{-1},
\label{eq:(3.8)}
\end{equation}
where
$R[\lambda(x)]_{\alpha\beta}=\delta_{\alpha\beta}\lambda(x,\mu)^{\mathrm{e}_\alpha}$
and
$R[\Lambda(x)]_{\alpha\beta}=\delta_{\alpha\beta}\Lambda(x,\mu)^{\mathrm{e}_\alpha'}$.

\subsection{Chirality projection and the fermion action}
\label{sec:3.2}
Now, the Ginsparg--Wilson relation~\eqref{eq:(3.5)} enables a clear separation
of chiralities of a lattice fermion. For this, we define the modified chiral
matrix by~\cite{Luscher:1998pqa,Niedermayer:1998bi},
\begin{equation}
   \Hat{\gamma}_5=\gamma_5(1-D).
\label{eq:(3.9)}
\end{equation}
Then, $(\Hat{\gamma}_5)^\dagger=\Hat{\gamma}_5$ and Eq.~\eqref{eq:(3.5)} implies
\begin{equation}
   (\Hat{\gamma}_5)^2=1,\qquad
   D\Hat{\gamma}_5=-\gamma_5D.
\label{eq:(3.10)}
\end{equation}
Thus, introducing chirality projection operators by
\begin{equation}
   \Hat{P}_\pm=\frac{1}{2}(1\pm\Hat{\gamma}_5),\qquad
   P_\pm=\frac{1}{2}(1\pm\gamma_5),
\label{eq:(3.11)}
\end{equation}
we may define the left-handed Weyl fermion on the lattice as
\begin{equation}
   \Hat{P}_-\psi=\psi,\qquad\Bar{\psi}P_+=\Bar{\psi}.
\label{eq:(3.12)}
\end{equation}
Then, the fermion action enjoys the form\footnote{\label{footnote:10}%
On~$\Gamma$, we assume the
periodic boundary conditions also on the fermion fields,
$\psi(x+L\Hat{\mu})=\psi(x)$. The overlap Dirac operator
in~Eq.~\eqref{eq:(3.6)} as it stands has an infinite range with an exponential
tail. The product in~Eq.~\eqref{eq:(3.13)} on the finite lattice~$\Gamma$ is
then defined by~$D\psi(x)=\sum_{y\in\Gamma}D_L(x,y)\psi(y)$,
where the kernel of the Dirac operator in finite volume is given by the
reflections, $D_L(x,y)=\sum_{n\in\mathbb{Z}^4}D(x,y+Ln)$.}
\begin{equation}
   S_{\mathrm{F}}=\sum_{x\in\Gamma}\Bar{\psi}(x)D\psi(x)
   =\sum_{x\in\Gamma}\Bar{\psi}(x)P_+D\Hat{P}_-\psi(x)
\label{eq:(3.13)}
\end{equation}
and chiralities are clearly separated in the lattice action (i.e., the action
of the Dirac fermion is given by the sum of this and the corresponding
right-handed one,
$S_{\mathrm{F}}=\sum_{x\in\Gamma}\Bar{\psi}(x)P_-D\Hat{P}_+\psi(x)$). This defines
the fermion action~$S_{\mathrm{F}}$ in~Eq.~\eqref{eq:(2.5)}.

The fermion integration measure in~Eq.~\eqref{eq:(2.5)}, on the other hand, is
defined by
\begin{equation}
   \mathrm{D}[\psi]\mathrm{D}[\Bar{\psi}]
   =\prod_j\mathrm{d}c_j\prod_k\mathrm{d}\Bar{c}_k
\label{eq:(3.14)}
\end{equation}
by employing (Grassmann-odd) expansion coefficients in
\begin{equation}
   \psi(x)=\sum_jv_j(x)c_j,\qquad
   \Bar{\psi}(x)=\sum_k\Bar{c}_k\Bar{v}_k(x),
\label{eq:(3.15)}
\end{equation}
where $v_j(x)$ and~$\Bar{v}_k(x)$ are orthonormal basis vectors in the
projected space~\eqref{eq:(3.12)}, i.e.
\begin{equation}
   \Hat{P}_-v_j=v_j,\qquad(v_k,v_j)=\delta_{kj}
\label{eq:(3.16)}
\end{equation}
and
\begin{equation}
   \Bar{v}_kP_-=\Bar{v}_k,
   \qquad(\Bar{v}_k^\dagger,\Bar{v}_j^\dagger)=\delta_{kj}.
\label{eq:(3.17)}
\end{equation}

Although the above provides a construction of a lattice Weyl fermion in the
classical level, this is just a beginning of the construction for the full
quantum theory~\cite{Luscher:1998du}. The point is that basis vectors~$v_j$
depend on the gauge field through the dependence of~$\Hat{P}_-$ on the gauge
field. The latter dependence, however, does not uniquely determine the basis
vectors; under a variation~$\delta$ of the gauge field,
from~Eq.~\eqref{eq:(3.16)},
\begin{equation}
   \Hat{P}_+\delta v_j=\delta\Hat{P}_-v_j.
\label{eq:(3.18)}
\end{equation}
Therefore, $\Hat{P}_-\delta v_j$ is not determined by~Eq.~\eqref{eq:(3.16)}
and there is wide ambiguity in the choice of basis
vectors~$v_j$.\footnote{Basis vectors $\Bar{v}_k$ can be chosen as being
independent of the gauge field and we will assume this in what follows.} One
has to choose the basis vectors so that physical requirements such as the
smoothness and locality of the fermion integration measure are
fulfilled~\cite{Luscher:1998du}.

We note that, in terms of the expansion coefficients in~Eq.~\eqref{eq:(3.15)},
the action~\eqref{eq:(3.13)} is written as
\begin{equation}
   S_{\mathrm{F}}
   =\sum_{k,j}\Bar{c}_kM_{kj}c_j,\qquad
   M_{kj}=\sum_{x\in\Gamma}\Bar{v}_k(x)Dv_j(x).
\label{eq:(3.19)}
\end{equation}
Corresponding to the fermion number anomaly in chiral gauge theory, the numbers
of $v_j$ and~$\Bar{v}_k$ can be generally different and if this is the case the
matrix~$M$ is rectangular. When the numbers of $v_j$ and~$\Bar{v}_k$ are the
same, $M$ is a square matrix and the partition function in the fermion sector
is given by
\begin{equation}
   \left\langle1\right\rangle_{\mathrm{F}}=\det M.
\label{eq:(3.20)}
\end{equation}
Also, in this case, the correlation functions of fermion fields are given by
the Wick contractions by the fermion propagator~$S_L(x,y)$,\footnote{$D_L(x,y)$
in this expression is the kernel of the overlap Dirac operator acting on
fermion fields in a finite volume; see footnote~\ref{footnote:10}.}
\begin{equation}
   \sum_{y\in\Gamma}D_L(x,y)S_L(y,z)=\delta_{xz}.
\label{eq:(3.21)}
\end{equation}
For instance, we have
\begin{equation}
   \left\langle\psi(x)\Bar{\psi}(y)\right\rangle_{\mathrm{F}}
   =\left\langle1\right\rangle_{\mathrm{F}}\times
   \Hat{P}_-S_L(x,y)P_+.
\label{eq:(3.22)}
\end{equation}

In what follows, we parametrize infinitesimal variations of the gauge fields as
\begin{equation}
   \delta_\eta u(x,\mu)=i\eta_\mu(x)u(x,\mu),\qquad
   \delta_\eta U(x,\mu)=i\eta_\mu'(x)U(x,\mu).
\label{eq:(3.23)}
\end{equation}
To parametrize variations of the basis vectors under these, we introduce the
measure term and measure currents by~\cite{Luscher:1998du},
\begin{equation}
   \mathfrak{L}_\eta=i\sum_j(v_j,\delta_\eta v_j)
   =:\sum_{x\in\Gamma}
   \left[\eta_\mu(x)j_\mu(x)+\eta_\mu'(x)J_\mu(x)\right].
\label{eq:(3.24)}
\end{equation}
Then the variation of the partition function~\eqref{eq:(3.20)} is given by
\begin{equation}
   \delta_\eta\ln\det M
   =\Tr_L(\delta_\eta D\Hat{P}_-D^{-1}P_+)-i\mathfrak{L}_\eta.
\label{eq:(3.25)}
\end{equation}

Particularly important variations are given by the gauge
transformations~\eqref{eq:(3.7)}, for which
\begin{equation}
   \eta_\mu(x)=-\partial_\mu\omega(x),\qquad
   \eta_\mu'(x)=-\partial_\mu\Omega(x),
\label{eq:(3.26)}
\end{equation}
where we have set $\lambda(x)=e^{i\omega(x)}$ and~$\Lambda(x)=e^{i\Omega(x)}$.
Under these, by the gauge covariance of the Dirac operator,
Eq.~\eqref{eq:(3.8)},
\begin{equation}
   \delta_\eta D=i[\omega t+\Omega T,D],
\label{eq:(3.27)}
\end{equation}
where the representation matrices are introduced by
\begin{equation}
   t_{\alpha\beta}=\delta_{\alpha\beta}\mathrm{e}_\alpha,\qquad
   T_{\alpha\beta}=\delta_{\alpha\beta}\mathrm{e}_\alpha'.
\label{eq:(3.28)}
\end{equation}
Then the gauge anomaly (i.e.\ noninvariance of the fermion partition
function~\eqref{eq:(3.20)} under the gauge transformation) is given
by\footnote{$\partial_\mu^*f(x)=f(x)-f(x-\Hat{\mu})$ is the backward difference
operator.}
\begin{equation}
   \delta_\eta\ln\det M
   =i\sum_{x\in\Gamma}\left\{\omega(x)\left[
   \mathcal{A}_L(x)-\partial_\mu^*j_\mu(x)
   \right]
   +\Omega(x)\left[
   \mathcal{A}_L'(x)-\partial_\mu^*J_\mu(x)
   \right]\right\},
\label{eq:(3.29)}
\end{equation}
where
\begin{equation}
   \mathcal{A}_L(x)=-\frac{1}{2}\tr\left[\gamma_5tD_L(x,x)\right],\qquad
   \mathcal{A}_L'(x)=-\frac{1}{2}\tr\left[\gamma_5TD_L(x,x)\right].
\label{eq:(3.30)}
\end{equation}

\subsection{Construction of the fermion integration measure}
\label{sec:3.3}
Appropriately modifying the proofs of the theorems
in~Ref.~\cite{Luscher:1998du}, we can show the following statements: Assuming
the admissibility~\eqref{eq:(2.10)} with a sufficiently small~$\epsilon<1/30$,
there exists a fermion integration measure which depends smoothly on gauge
fields. The variation of the measure under the change of the dynamical gauge
field~$u(x,\mu)$ depends locally on~$u(x,\mu)$ but nonlocally on the external
gauge field~$U(x,\mu)$. The variation of the measure under the change of the
external gauge field~$U(x,\mu)$ depends nonlocally on the dynamical gauge
field~$u(x,\mu)$. These nonlocal dependences are worrisome, but we do not
think they are problematic, because $U(x,\mu)$ is not dynamical and is not the
subject of the functional integral. The fermion measure possesses the gauge
transformation property under the infinitesimal gauge
transformations~\eqref{eq:(3.26)}, as
\begin{equation}
   \delta_\eta\langle\mathcal{O}\rangle_{\mathrm{F}}
   =\langle\delta_\eta\mathcal{O}\rangle_{\mathrm{F}}
   -2i\gamma
   \sum_{x\in\Gamma}\Omega(x)
   \epsilon_{\mu\nu\rho\sigma}
   f_{\mu\nu}(x)f_{\rho\sigma}(x+\Hat{\mu}+\Hat{\nu})
   \langle\mathcal{O}\rangle_{\mathrm{F}},
\label{eq:(3.31)}
\end{equation}
where:\footnote{Our convention is
$\gamma_5=\gamma_0\gamma_1\gamma_2\gamma_3$ with Hermitian Dirac
matrices~$\gamma_\mu$ and $\epsilon_{0123}=1$. In particular,
$\tr(\gamma_5\gamma_\mu\gamma_\nu\gamma_\rho\gamma_\sigma)=%
4\epsilon_{\mu\nu\rho\sigma}$.}
\begin{equation}
   \gamma=-\frac{1}{32\pi^2}.
\label{eq:(3.32)}
\end{equation}
In the expectation value~\eqref{eq:(3.31)}, gauge transformations on the
fermion fields are given by
\begin{equation}
   \delta_\eta\psi(x)=\left[i\omega(x)t+i\Omega(x)T\right]\psi(x),\qquad
   \delta_\eta\Bar{\psi}(x)
   =\Bar{\psi}(x)\left[-i\omega(x)t-i\Omega(x)T\right].
\label{eq:(3.33)}
\end{equation}
Equation~\eqref{eq:(3.31)} shows that the first $U(1)$ has no gauge anomaly
whereas the second $U(1)'$ possesses the gauge anomaly with a definite
structure. The anomaly relation~\eqref{eq:(3.31)} is the most fundamental
relation for our construction of the symmetry operator in the next section.

The proof of the above statements proceeds as follows.

First, the above statements follow if there exist currents $j_\mu(x)$
and~$J_\mu(x)$ which fulfill the following conditions: (i)~The currents depend
smoothly on gauge fields and $j_\mu(x)$ depends locally on the dynamical gauge
field~$u(x,\mu)$. (ii)~The functional
\begin{equation}
   \mathfrak{L}_\eta
   =\sum_{x\in\Gamma}
   \left[\eta_\mu(x)j_\mu(x)+\eta_\mu'(x)J_\mu(x)\right]
\label{eq:(3.34)}
\end{equation}
satisfies the ``integrability condition,''\footnote{Here, $\Tr_L$ is the trace
over the fermion fields on~$\Gamma$.}
\begin{equation}
   \delta_\eta\mathfrak{L}_\zeta
   -\delta_\zeta\mathfrak{L}_\eta
   =i\Tr_L(\Hat{P}_-[\delta_\eta\Hat{P}_-,\delta_\zeta\Hat{P}_-]).
\label{eq:(3.35)}
\end{equation}
(iii)~ The currents satisfies the ``anomalous conservation laws,''
\begin{equation}
   \partial_\mu^*j_\mu(x)=\mathcal{A}_L(x),\qquad
   \partial_\mu^*J_\mu(x)=\mathcal{A}_L'(x)
   +2\gamma\epsilon_{\mu\nu\rho\sigma}
   f_{\mu\nu}(x)f_{\rho\sigma}(x+\Hat{\mu}+\Hat{\nu}).
\label{eq:(3.36)}
\end{equation}

This part of the proof corresponds to the proof of Theorem~5.1
of~Ref.~\cite{Luscher:1998du} (Sect.~10). The idea is
that~\cite{Luscher:1998du} the measure term~$\mathfrak{L}_\eta$
in~Eq.~\eqref{eq:(3.24)} can be regarded as a connection of a principal $U(1)$
bundle over the configuration space of admissible gauge fields, which is
denoted by~$\mathfrak{U}$ in what follows. The curvature 2-form of the bundle
is given by the quantity on the right-hand side of~Eq.~\eqref{eq:(3.35)}. The
global existence of the currents $j_\mu(x)$ and~$J_\mu(x)$ on~$\mathfrak{U}$
and the corresponding functional~$\mathfrak{L}_\eta$ which
satisfies~Eq.~\eqref{eq:(3.35)} then implies that this $U(1)$ bundle is
trivial. Writing a measure term defined from a certain set of basis
vectors~$v_j$ by
\begin{equation}
   \widetilde{\mathfrak{L}}_\eta=i\sum_j(v_j,\delta_\eta v_j),
\label{eq:(3.37)}
\end{equation}
there exists a fermion integration measure with the above properties if and
only if, for any closed curve in~$\mathfrak{U}$,
\begin{equation}
   u_s(x,\mu),\qquad U_s(x,\mu),\qquad0\leq s\leq2\pi,
\label{eq:(3.38)}
\end{equation}
the Wilson lines
\begin{equation}
   \mathfrak{W}=\exp\left(i\int_0^{2\pi}ds\,\mathfrak{L}_\eta\right),\qquad
   \widetilde{\mathfrak{W}}
   =\exp\left(i\int_0^{2\pi}ds\,\widetilde{\mathfrak{L}}_\eta\right),
\label{eq:(3.39)}
\end{equation}
where
$\eta_\mu(x)=%
-iu_s(x,\mu)^{-1}\partial_s u_s(x,\mu)-iU_s(x,\mu)^{-1}\partial_s U_s(x,\mu)$,
coincide, i.e.\ $\mathfrak{W}=\widetilde{\mathfrak{W}}$.

To show this, one has to compute the above Wilson lines for all nontrivial
loops in~$\mathfrak{U}$. In~Ref.~\cite{Luscher:1998du}, it is shown that
$\mathfrak{U}$ in $U(1)$ lattice gauge theory has the topology of a
multidimensional torus~$T^n=S^1\times S^1\times\dotsb\times S^1$ times a
contractive space (here, four $S^1$'s correspond to $U(1)$ Wilson loops in
real space and the others correspond to $U(1)$ gauge transformations). Thus, it
suffices to consider Wilson lines wrapping each $S^1$. In our present system,
as far as the pure-gauge external field~\eqref{eq:(2.4)} without integration
over~$\Lambda(x)$ with nontrivial windings (this is the situation we are
interested in) is considered, no new $S^1$ in the $U(1)'$ gauge theory
emerges.\footnote{Actually, from the computation in~Sect.~13.3
of~Ref.~\cite{Luscher:1998du}, we see that it is impossible to construct a
smooth fermion integration measure along the gauge loop in the $U(1)'$
direction because of the gauge anomaly.} Therefore, we may invoke the results
of~Sect.~10 of~Ref.~\cite{Luscher:1998du} without change.

Next, we have to show the existence of the currents in~Eq.~\eqref{eq:(3.34)}
with required properties. As per Theorem~5.3 of~Ref.~\cite{Luscher:1998du}, we
first show the existence of currents with similar properties in an infinite
volume lattice, $\mathbb{Z}^4$. That is, we show that there exist currents
$j_\mu^\star(x)$ and~$J_\mu^\star(x)$ on~$\mathbb{Z}^4$ which fulfill the
following conditions: (i) Both currents depend smoothly on gauge fields
and $j_\mu^\star(x)$ depends locally on~$u(x,\mu)$. (ii)~The corresponding
functional,\footnote{On infinite volume, the vector fields $\eta_\mu(x)$
and~$\eta_\mu'(x)$ are supposed to have finite supports on~$\mathbb{Z}^4$.}
\begin{equation}
   \mathfrak{L}_\eta^\star=\sum_{x\in\mathbb{Z}^4}
   \left[\eta_\mu(x)j_\mu^\star(x)+\eta_\mu'(x)J_\mu^\star(x)\right],
\label{eq:(3.40)}
\end{equation}
satisfies the integrability condition,
\begin{equation}
   \delta_\eta\mathfrak{L}_\zeta^\star
   -\delta_\zeta\mathfrak{L}_\eta^\star
   =i\Tr(\Hat{P}_-[\delta_\eta\Hat{P}_-,\delta_\zeta\Hat{P}_-]).
\label{eq:(3.41)}
\end{equation}
(iii)~The currents satisfy the anomalous conservation laws,
\begin{equation}
   \partial_\mu^*j_\mu^\star(x)=\mathcal{A}(x),\qquad
   \partial_\mu^*J_\mu^\star(x)=\mathcal{A}'(x)
   +2\gamma\epsilon_{\mu\nu\rho\sigma}
   f_{\mu\nu}(x)f_{\rho\sigma}(x+\Hat{\mu}+\Hat{\nu}),
\label{eq:(3.42)}
\end{equation}
where
\begin{equation}
   \mathcal{A}(x)=-\frac{1}{2}\tr\left[\gamma_5tD(x,x)\right],\qquad
   \mathcal{A}'(x)=-\frac{1}{2}\tr\left[\gamma_5TD(x,x)\right].
\label{eq:(3.43)}
\end{equation}
Note that all these expressions refer to an infinite volume.

In an infinite volume, one can construct the required currents rather
explicitly. The point is that, in infinite volume, one can parametrize
admissible gauge fields by vector fields as~\cite{Luscher:1998kn},
\begin{align}
   u(x,\mu)&=\mathrm{e}^{i\mathfrak{a}_\mu(x)},
   &|\mathfrak{a}_\mu(x)|&\leq\pi(1+8\|x\|),&
   f_{\mu\nu}(x)&=\partial_\mu\mathfrak{a}_\nu(x)
   -\partial_\nu\mathfrak{a}_\mu(x),
\notag\\
   U(x,\mu)&=\mathrm{e}^{i\mathfrak{A}_\mu(x)},
   &|\mathfrak{A}_\mu(x)|&\leq\pi(1+8\|x\|),&
   F_{\mu\nu}(x)&=\partial_\mu\mathfrak{A}_\nu(x)
   -\partial_\nu\mathfrak{A}_\mu(x).
\label{eq:(3.44)}
\end{align}
This is shown in the following way~\cite{Luscher:1998kn,Hernandez:1995jc}. We
recall Eq.~\eqref{eq:(2.13)} which holds for admissible gauge fields on
infinite volume as well as on~$\Gamma$. Then, we can solve
\begin{equation}
   \partial_\mu z_\nu(x)-\partial_\nu z_\mu(x)=z_{\mu\nu}(x)
\label{eq:(3.45)}
\end{equation}
on~$\mathbb{Z}^4$ by taking the complete axial gauge referring to a certain
point~$o$ in~$\mathbb{Z}^4$. Then, $\mathfrak{a}_\mu(x)$ is given by 
\begin{equation}
   \mathfrak{a}_\mu(x)=a_\mu(x)+2\pi z_\mu(x).
\label{eq:(3.46)}
\end{equation}
The important point here is that the definition of $\mathfrak{a}_\mu(x)$ refers
to the point~$o$ and it depends on the choice of~$o$. However, it can be seen
that different choice of~$o$ corresponds to a different choice of the gauge and
the dependence on~$o$ disappears in gauge invariant quantities. This fact plays
a crucial role in the construction in~Ref.~\cite{Luscher:1998du}.

Now, using the parametrizations in~Eq.~\eqref{eq:(3.44)}, from the topological
invariance of~$\mathcal{A}(x)$ and~$\mathcal{A}'(x)$,\footnote{Taking a local
variation of~$(\Hat{\gamma}_5)^2=1$ in~Eq.~\eqref{eq:(3.10)}, we have
$\Hat{\gamma}_5\gamma_5\delta_\eta D=-\gamma_5\delta_\eta D\Hat{\gamma}_5$.
Using this, $\sum_{x\in\mathbb{Z}^4}\delta_\eta\mathcal{A}(x)
=(-1/2)\Tr(\gamma_5t\delta_\eta D)
=(-1/2)\Tr[(\Hat{\gamma}_5)^2\gamma_5t\delta_\eta D]=0$ follows; similarly, we
have $\sum_{x\in\mathbb{Z}^4}\delta_\eta\mathcal{A}'(x)=0$. These are the
topological invariance of~$\mathcal{A}(x)$ and~$\mathcal{A}'(x)$ meant in the
text.} an ingenious cohomological argument with finite lattice spacings
shows~\cite{Luscher:1998kn},
\begin{equation}
   \mathcal{A}(x)=\partial_\mu^*k_\mu(x),\qquad
   \mathcal{A}'(x)
   =-2\gamma\epsilon_{\mu\nu\rho\sigma}
   f_{\mu\nu}(x)f_{\rho\sigma}(x+\Hat{\mu}+\Hat{\nu})
   +\partial_\mu^*K_\mu(x),
\label{eq:(3.47)}
\end{equation}
where $\gamma$ is a constant\footnote{We will later argue that this constant is
given by~Eq.~\eqref{eq:(3.32)}.} and the currents $k_\mu(x)$ and~$K_\mu(x)$ are
gauge invariant. See also~Ref.~\cite{Fujiwara:1999fi}. Here, we have used the
charge assignment in~Eq.~\eqref{eq:(3.1)} to determine possible anomalous terms
of the form~$\sim\epsilon_{\mu\nu\rho\sigma}f_{\mu\nu}f_{\rho\sigma}$. Generally
speaking, we may also have anomalous terms of the forms, $\mathcal{A}(x)\sim
\epsilon_{\mu\nu\rho\sigma}f_{\mu\nu}F_{\rho\sigma}$
and~$\mathcal{A}'(x)\sim\epsilon_{\mu\nu\rho\sigma}F_{\mu\nu}F_{\rho\sigma}$.
However, since $F_{\mu\nu}(x)=0$ in our problem as noted
in~Eq.~\eqref{eq:(2.17)}, we do not write these possible anomalies
in~Eq.~\eqref{eq:(3.47)}.

With these preparations, we define the measure term in infinite volume by
\begin{equation}
   \mathfrak{L}_\eta^\star
   =\mathfrak{L}_\eta^{\star\text{inv}}
   +\mathfrak{L}_\eta^{\star\text{non-inv}}
\label{eq:(3.48)}
\end{equation}
where
\begin{align}
   \mathfrak{L}_\eta^{\star\text{inv}}
   &=i\int_0^1ds\,
   \Tr(\Hat{P}_-[\partial_s\Hat{P}_-,\delta_\eta\Hat{P}_-])
   +\int_0^1ds\,\sum_{x\in\mathbb{Z}^4}
   \left[
   \eta_\mu(x)k_\mu(x)+\mathfrak{a}_\mu(x)\delta_\eta k_\mu(x)
   \right]
\notag\\
   &\qquad{}
   +\int_0^1ds\,\sum_{x\in\mathbb{Z}^4}
   \left[
   \eta_\mu'(x)K_\mu(x)+\mathfrak{A}_\mu(x)\delta_\eta K_\mu(x)
   \right]
\notag\\
   &\qquad{}
   -\frac{4}{3}\gamma
   \sum_{x\in\mathbb{Z}^4}\epsilon_{\mu\nu\rho\sigma}
   \bigr\{
   \eta_\mu'(x)\mathfrak{a}_\nu(x+\Hat{\mu})f_{\rho\sigma}(x+\Hat{\mu}+\Hat{\nu})
\notag\\
   &\qquad\qquad\qquad\qquad\qquad{}
   +\mathfrak{A}_\mu(x)\delta_\eta
   \left[\mathfrak{a}_\nu(x+\Hat{\mu})f_{\rho\sigma}(x+\Hat{\mu}+\Hat{\nu})\right]
   \bigr\}
\notag\\
   &=:\sum_{x\in\mathbb{Z}^4}
   \left[\eta_\mu(x)j_\mu^{\star\text{inv}}(x)
   +\eta_\mu'(x)J_\mu^{\star\text{inv}}(x)\right],
\label{eq:(3.49)}
\end{align}
and
\begin{align}
   \mathfrak{L}_\eta^{\star\text{non-inv}}
   &=4\gamma
   \sum_{x\in\mathbb{Z}^4}\epsilon_{\mu\nu\rho\sigma}
   \bigr\{
   \eta_\mu'(x)\mathfrak{a}_\nu(x+\Hat{\mu})f_{\rho\sigma}(x+\Hat{\mu}+\Hat{\nu})
\notag\\
   &\qquad\qquad\qquad\qquad{}
   +\mathfrak{A}_\mu(x)\delta_\eta
   \left[
   \mathfrak{a}_\nu(x+\Hat{\mu})f_{\rho\sigma}(x+\Hat{\mu}+\Hat{\nu})\right]
   \bigr\}
\notag\\
   &=:\sum_{x\in\mathbb{Z}^4}
   \left[\eta_\mu(x)j_\mu^{\star\text{non-inv}}(x)
   +\eta_\mu'(x)J_\mu^{\star\text{non-inv}}(x)\right].
\label{eq:(3.50)}
\end{align}
In Eq.~\eqref{eq:(3.49)}, the dependence of gauge fields on the parameter~$s$
is introduced using the parametrization in~Eq.~\eqref{eq:(3.44)} as
\begin{equation}
   u_s(x,\mu)=\mathrm{e}^{is\mathfrak{a}_\mu(x)},\qquad
   U_s(x,\mu)=\mathrm{e}^{is\mathfrak{A}_\mu(x)},\qquad
   0\leq s\leq1.
\label{eq:(3.51)}
\end{equation}

The smooth dependence of~$\mathfrak{L}_\eta^\star$~\eqref{eq:(3.48)} on gauge
fields is obvious from the well-definedness of the overlap Dirac
operator for admissible gauge fields
with~$\epsilon<1/30$~\cite{Hernandez:1998et}. As per Sect.~6
of~Ref.~\cite{Luscher:1998du}, the integrability condition~\eqref{eq:(3.41)}
can be shown straightforwardly for the sum of expressions
in~Eqs.~\eqref{eq:(3.49)} and~\eqref{eq:(3.50)}. In particular, only the first
term on the right-hand side of~Eq.~\eqref{eq:(3.49)} contributes to the
integrability condition, because the other terms in~Eqs.~\eqref{eq:(3.49)}
and~\eqref{eq:(3.50)} are $\delta_\eta$ of certain combinations. The sum of
terms in~Eqs.~\eqref{eq:(3.49)} and~\eqref{eq:(3.50)} being proportional
to~$\gamma$ is peculiar to our ``anomalous gauge theory'' and it is a variation
of a counterterm,
$-i\Delta=(8/3)\gamma\sum_{x\in\mathbb{Z}^4}\epsilon_{\mu\nu\rho\sigma}
\mathfrak{A}_\mu(x)\mathfrak{a}_\nu(x+\Hat{\mu})
f_{\rho\sigma}(x+\Hat{\mu}+\Hat{\nu})$. This counterterm makes the vector current
in the target theory conserving and the coefficient of the
$\epsilon_{\mu\nu\rho\sigma}f_{\mu\nu}f_{\rho\sigma}$ term in the axial anomaly the
naively expected one. See Appendix~\ref{sec:A} for a detailed exposition on
this point.

To show Eq.~\eqref{eq:(3.42)}, we set $\eta_\mu(x)=-\partial_\mu\omega(x)$
and~$\eta_\mu'(x)=-\partial_\mu\Omega(x)$ in~Eqs.~\eqref{eq:(3.48)}. The
left-hand side then becomes
\begin{equation}
   \sum_{x\in\mathbb{Z}^4}
   \left[\omega(x)\partial_\mu^*j_\mu^\star(x)
   +\Omega(x)\partial_\mu^*J_\mu^\star(x)\right].
\label{eq:(3.52)}
\end{equation}
On the other hand, noting the gauge covariance
$\delta_\eta\Hat{P}_-=is[\omega t+\Omega T,\Hat{P}_-]$ (this follows
from~Eq.~\eqref{eq:(3.27)}), the gauge invariance of~$k_\mu(x)$ and~$K_\mu(x)$,
and the identity $\Hat{P}_-\delta_\eta\Hat{P}_-\Hat{P}_-=0$,
from~Eqs.~\eqref{eq:(3.49)} and~\eqref{eq:(3.50)}, we find that
\begin{align}
   \mathfrak{L}_\eta^{\star\text{inv}}
   &=
   \sum_{x\in\mathbb{Z}^4}
   \left.\left[\omega(x)\mathcal{A}(x)
   +\Omega(x)\mathcal{A}(x)\right]\right|_{s=1},
\notag\\
   \mathfrak{L}_\eta^{\star\text{non-inv}}
   &=2\gamma\sum_{x\in\mathbb{Z}^4}
   \Omega(x)\epsilon_{\mu\nu\rho\sigma}
   f_{\mu\nu}(x)f_{\rho\sigma}(x+\Hat{\mu}+\Hat{\nu}),
\label{eq:(3.53)}
\end{align}
where we have used Eqs.~\eqref{eq:(3.43)} and~\eqref{eq:(3.47)}; we have also
used the Bianchi identity~\eqref{eq:(2.16)} and $F_{\mu\nu}(x)=0$. This shows
Eq.~\eqref{eq:(3.42)}.

An important difference of our system from that of~Ref.~\cite{Luscher:1998du}
is the gauge noninvariance of the currents, $j_\mu^\star(x)$
and~$J_\mu^\star(x)$. Under the gauge transformations,
$\mathfrak{a}_\mu(x)\to\mathfrak{a}_\mu(x)+\partial_\mu\omega(x)$
and~$\mathfrak{A}_\mu(x)\to\mathfrak{A}_\mu(x)+\partial_\mu\Omega(x)$, in a
similar way to deriving Eq.~\eqref{eq:(3.53)}, we find
\begin{align}
   \mathfrak{L}_\eta^{\star\text{inv}}
   &\to\mathfrak{L}_\eta^{\star\text{inv}},
\notag\\
   \mathfrak{L}_\eta^{\star\text{non-inv}}
   &\to\mathfrak{L}_\eta^{\star\text{non-inv}}
   -2\gamma\sum_{x\in\mathbb{Z}^4}\Omega(x)
   \delta_\eta\epsilon_{\mu\nu\rho\sigma}
   f_{\mu\nu}(x)f_{\rho\sigma}(x+\Hat{\mu}+\Hat{\nu})
\notag\\
   &\qquad\qquad\qquad{}
   +\frac{8}{3}\gamma\sum_{x\in\mathbb{Z}^4}
   \epsilon_{\mu\nu\rho\sigma}\eta_\mu'(x)\partial_\nu\omega(x+\Hat{\mu})
   f_{\rho\sigma}(x+\Hat{\mu}+\Hat{\nu}).
\label{eq:(3.54)}
\end{align}
This shows that the current $j_\mu^\star(x)$ in~Eq.~\eqref{eq:(3.40)} is not
invariant under the $U(1)'$ gauge transformation and the
current~$J_\mu^\star(x)$ in~Eq.~\eqref{eq:(3.40)} is not invariant under the
$U(1)$ gauge transformation. These gauge noninvariances have an important
consequence on the locality of the currents, because the relation between the
vector fields $\mathfrak{a}_\mu(x)$ and~$\mathfrak{A}_\mu(x)$
in~Eq.~\eqref{eq:(3.44)}, from which the measure term~\eqref{eq:(3.48)} is
constructed, and link variables $u(x,\mu)$ and~$U(x,\mu)$ is local only up to
the $U(1)\times U(1)'$ gauge transformations~\cite{Luscher:1998kn}. A closer
look at~Eq.~\eqref{eq:(3.54)} shows that $J_\mu^{\star\text{non-inv}}(x)$, which
couples to the external gauge field~$U(x,\mu)$, does not locally depend on the
dynamical gauge field~$u(x,\mu)$ and $j_\mu^{\star\text{non-inv}}(x)$, which
couples to the dynamical gauge field~$u(x,\mu)$, does not locally depend on the
external gauge field~$U(x,\mu)$. However, since the gauge field $U(x,\mu)$ is
not dynamical, we do not think these nonlocalities are problematic.

We have to then show the existence of the currents $j_\mu(x)$ and~$J_\mu(x)$
on the finite volume lattice~$\Gamma$, which possess the properties (i)--(iii)
listed around~Eq.~\eqref{eq:(3.34)}. This is accomplished~\cite{Luscher:1998du}
by starting from the currents $j_\mu^\star(x)$ and~$J_\mu^\star(x)$ in infinite
volume and invoking the locality of the overlap Dirac operator~$D$, which is
guaranteed for~$\epsilon<1/30$~\cite{Hernandez:1998et}. First, setting
\begin{equation}
   \mathfrak{L}_\eta^{\text{inv}}
   =\sum_{x\in\Gamma}
   \left[\eta_\mu(x)j_\mu^{\text{inv}}(x)
   +\eta_\mu'(x)J_\mu^{\text{inv}}(x)\right],
\label{eq:(3.55)}
\end{equation}
we can show the existence of~$j_\mu^{\text{inv}}(x)$ and~$J_\mu^{\text{inv}}(x)$
such that
\begin{equation}
   \delta_\eta\mathfrak{L}_\zeta^{\text{inv}}
   -\delta_\zeta\mathfrak{L}_\eta^{\text{inv}}
   =i\Tr_L(\Hat{P}_-[\delta_\eta\Hat{P}_-,\delta_\zeta\Hat{P}_-])
\label{eq:(3.56)}
\end{equation}
and
\begin{equation}
   \partial_\mu^*j_\mu^{\text{inv}}(x)=\mathcal{A}_L(x),\qquad
   \partial_\mu^*J_\mu^{\text{inv}}(x)=\mathcal{A}_L'(x).
\label{eq:(3.57)}
\end{equation}
This is done by starting from the currents $j_\mu^{\star\text{inv}}(x)$
and~$J_\mu^{\star\text{inv}}(x)$ in infinite volume given
by~Eq.~\eqref{eq:(3.49)} and repeating the argument in~Sect.~11
of~Ref.~\cite{Luscher:1998du}. The measure term
$\mathfrak{L}_\eta^{\star\text{inv}}$~\eqref{eq:(3.49)} fulfills the same
prerequisites as the measure term~$\mathfrak{L}_\eta^\star$
of~Ref.~\cite{Luscher:1998du}. In particular, since the currents
$j_\mu^{\star\text{inv}}(x)$ and~$J_\mu^{\star\text{inv}}(x)$ are gauge invariant
as~Eq.~\eqref{eq:(3.54)} shows, $j_\mu^{\star\text{inv}}(x)$
and~$J_\mu^{\star\text{inv}}(x)$ locally depend on gauge fields. Thus, we can
literally apply the argument of~Ref.~\cite{Luscher:1998du} to show the
existence of~$j_\mu^{\text{inv}}(x)$ and~$J_\mu^{\text{inv}}(x)$ on~$\Gamma$.

The currents $j_\mu^{\text{inv}}(x)$ and~$J_\mu^{\text{inv}}(x)$ obtained above,
however, are not quite the same as the desired currents $j_\mu(x)$
and~$J_\mu(x)$. A comparison of Eqs.~\eqref{eq:(3.57)} and~\eqref{eq:(3.36)}
shows that $J_\mu^{\text{inv}}(x)$ does not have the desired anomaly. To remedy
this point, we define
\begin{align}
   \mathfrak{L}_\eta^{\text{non-inv}}
   &=\sum_{x\in\Gamma}
   \left[\eta_\mu(x)j_\mu^{\text{non-inv}}(x)
   +\eta_\mu'(x)J_\mu^{\text{non-inv}}(x)\right]
\notag\\
   &=4\gamma
   \sum_{x\in\Gamma}\epsilon_{\mu\nu\rho\sigma}
   \bigr\{
   \eta_\mu'(x)\mathfrak{a}_\nu(x+\Hat{\mu})f_{\rho\sigma}(x+\Hat{\mu}+\Hat{\nu})
\notag\\
   &\qquad\qquad\qquad\qquad{}
   +\mathfrak{A}_\mu(x)\delta_\eta
   \left[
   \mathfrak{a}_\nu(x+\Hat{\mu})f_{\rho\sigma}(x+\Hat{\mu}+\Hat{\nu})\right]
   \bigr\}.
\label{eq:(3.58)}
\end{align}
In this expression, the gauge potentials $\mathfrak{a}_\mu(x)$
and~$\mathfrak{A}_\mu(x)$ are given by first extending the gauge fields
on~$\Gamma$ to $\mathbb{Z}^4$ by periodic copies and then apply the
construction as~Eq.~\eqref{eq:(3.44)}.\footnote{From this description, one may
wonder if this construction can be used to directly define the Chern--Simons
form
as~$\epsilon_{\mu\nu\rho\sigma}\mathfrak{a}_\nu(x+\Hat{\mu})f_{\rho\sigma}(x+\Hat{\mu}+\Hat{\nu})$. We do not pursue this possibility here, because we are
interested in how the axial anomaly arising from the fermion can be captured in
a lattice framework.} Then, since the currents $j_\mu^{\text{non-inv}}(x)$
and~$J_\mu^{\text{non-inv}}(x)$ are not gauge invariant, these currents generally
do not depend locally on gauge fields. As in infinite volume,
$J_\mu^{\text{non-inv}}(x)$, which couples to the external gauge
field~$U(x,\mu)$, does not locally depend on the dynamical gauge
field~$u(x,\mu)$ and $j_\mu^{\text{non-inv}}(x)$, which couples to the dynamical
gauge field~$u(x,\mu)$, does not locally depend on the external gauge
field~$U(x,\mu)$. We do not think, however, these nonlocalities are harmful
because $U(x,\mu)$ is an external field and is not dynamical.

Finally, we define the full measure currents by
\begin{equation}
   j_\mu(x)
   =j_\mu^{\text{inv}}(x)+j_\mu^{\text{non-inv}}(x),\qquad
   J_\mu(x)
   =J_\mu^{\text{inv}}(x)+J_\mu^{\text{non-inv}}(x).
\label{eq:(3.59)}
\end{equation}
The addition of $j_\mu^{\text{non-inv}}(x)$ and~$J_\mu^{\text{non-inv}}(x)$ does
not influences the integrability~\eqref{eq:(3.56)}, because 
$\mathfrak{L}_\eta^{\text{non-inv}}$~\eqref{eq:(3.58)} is $\delta_\eta$ of a
certain combination. The desired anomaly of~$J_\mu(x)$~\eqref{eq:(3.36)} is
reproduced by~$J_\mu^{\text{non-inv}}(x)$ as in~Eq.~\eqref{eq:(3.53)}. This
completes our argument on the existence of the fermion integration
measure.\footnote{Although here we have imitated the construction
of~Ref.~\cite{Luscher:1998du} that starts with the cohomological argument in
infinite volume~\cite{Luscher:1998kn} and then relates it to the construction
in finite volume, it might be possible to imitate the argument
of~Ref.~\cite{Kadoh:2007wz} that directly constructs the fermion measure in
finite volume.}

At the end of this section, we argue that the value of the
coefficient~$\gamma$ in the above expressions is given
by~Eq.~\eqref{eq:(3.32)}. First, one can compute the
anomaly~$\mathcal{A}'(x)$~\eqref{eq:(3.43)} in the classical continuum
limit~\cite{Kikukawa:1998pd,Fujikawa:1998if,Suzuki:1998yz,Adams:1998eg}. The
comparison of the answer (when $D$ does not possess the species doublers)
with~Eq.~\eqref{eq:(3.47)} gives Eq.~\eqref{eq:(3.32)} (the factor~$-2$ comes
from~$\tr(T)=-2$). This might not be quite satisfactory, because one invokes the
classical continuum limit. Another argument is based on the lattice index
theorem~\cite{Neuberger:1997fp,Hasenfratz:1997ft,Hasenfratz:1998ri,Neuberger:1998wv,Hasenfratz:1998jp,Luscher:1998pqa,Niedermayer:1998bi}. Let us
consider a single Dirac fermion with the $U(1)$ charge~$+1$. The
Ginsparg--Wilson relation~\eqref{eq:(3.5)} can be written
as~$\{H,{\mit\Gamma}_5\}=0$, where $H=\gamma_5D$ is the Hermitian Dirac
operator and ${\mit\Gamma}_5:=\gamma_5(1-D/2)=\gamma_5-H/2$. This shows that,
for an eigenvector~$u_n$ for which $Hu_n=\lambda_nu_n$, ${\mit\Gamma}_5u_n$
possesses the opposite eigenvalue~$-\lambda_n$. Therefore, $u_n$
and~${\mit\Gamma}_5u_n$ are orthogonal if~$\lambda_n\neq0$. From this, assuming
that $u_n$ is normalized,
\begin{equation}
   \Tr_L{\mit\Gamma}_5=\sum_n(u_n,{\mit\Gamma}_5u_n)=n_+-n_-,
\label{eq:(3.60)}
\end{equation}
where $n_\pm$ are the numbers of zero modes, $Hu_n=0$, with positive and
negative chiralities, ${\mit\Gamma}_5=\gamma_5=\pm1$, respectively. We note
that the left-hand side of~Eq.~\eqref{eq:(3.60)} can also be written as
\begin{align}
   \sum_{x\in\Gamma}\tr\left[{\mit\Gamma}_5(x,x)\right]
   &=-\frac{1}{2}\sum_{x\in\Gamma}\tr\left[\gamma_5D(x,x)\right]
\notag\\
   &=\gamma\sum_{x\in\Gamma}\epsilon_{\mu\nu\rho\sigma}
   f_{\mu\nu}(x)f_{\rho\sigma}(x+\Hat{\mu}+\Hat{\nu})
\notag\\
   &=4\pi^2\gamma\epsilon_{\mu\nu\rho\sigma}
   m_{\mu\nu}m_{\rho\sigma},
\label{eq:(3.61)}
\end{align}
where, in the second equality, we have used~\cite{Luscher:1998kn},
\begin{equation}
   -\frac{1}{2}\tr\left[\gamma_5D(x,x)\right]
   =\gamma\epsilon_{\mu\nu\rho\sigma}
   f_{\mu\nu}(x)f_{\rho\sigma}(x+\Hat{\mu}+\Hat{\nu})
   +\partial_\mu^*k_\mu(x)
\label{eq:(3.62)}
\end{equation}
with $k_\mu(x)$ being a gauge invariant current
on~$\mathbb{Z}^4$.\footnote{This is different from the current~$k_\mu(x)$
defined in~Eq.~\eqref{eq:(3.47)}.} The gauge field in this expression is given
by periodic copies of that on~$\Gamma$. Since $k_\mu(x)$ is gauge invariant, it
is periodic on~$\Gamma$ and thus $\sum_{x\in\Gamma}\partial_\mu^*k_\mu(x)=0$.
In the last equality, we have used Eq.~(7.14)~of~Ref.~\cite{Luscher:1998du};
the integers $m_{\mu\nu}$ are the magnetic fluxes defined
by~Eq.~\eqref{eq:(2.11)}. The equality of Eqs.~\eqref{eq:(3.60)}
and~\eqref{eq:(3.61)},
$n_+-n_-=4\pi^2\gamma\epsilon_{\mu\nu\rho\sigma}m_{\mu\nu}m_{\rho\sigma}$,
is the lattice index theorem in Abelian gauge theory. Since the minimal value
of~$\epsilon_{\mu\nu\rho\sigma}m_{\mu\nu}m_{\rho\sigma}$ is~$8$, the
coefficient~\eqref{eq:(3.32)} is consistent with this index theorem. From these
facts, we believe that the value in~Eq.~\eqref{eq:(3.32)} is the only
possible one for a physically sensible lattice Dirac operator.\footnote{It
appears that it is possible to give a mathematical proof on the value of this
coefficient~\cite{Fukaya:2023xx}.}

\section{Symmetry operator and the topological defect for the axial $U(1)$
noninvertible symmetry}
\label{sec:4}
\subsection{Introduction of the topological defect}
\label{sec:4.1}
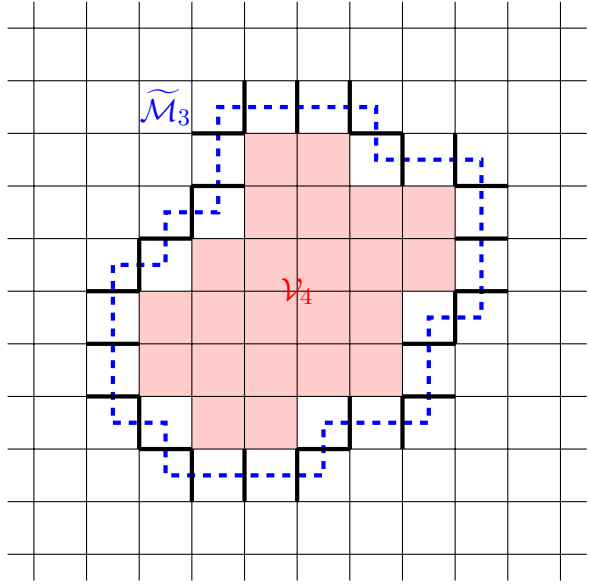
\begin{figure}[htbp]
\centering
\begin{tikzpicture}[scale=0.7]
  \fill[red!20]
 (2,3)--
 (3,3)--
 (3,2)--
 (4,2)--
 (4,2)--
 (5,2)--
 (5,3)--
 (7,3)--
 (7,5)--
 (8,5)--
 (8,7)--
 (6,7)--
 (6,8)--
 (4,8)--
 (4,6)--
 (3,6)--
 (3,5)--
 (2,5)--
 (2,3);
  \draw (-0.5,-0.5) grid[step=1] (10.5,10.5);
  \draw[dashed,blue,ultra thick]
 (1.5,2.5)--
 (2.5,2.5)--
 (2.5,1.5)--
 (3.5,1.5)--
 (3.5,1.5)--
 (5.5,1.5)--
 (5.5,2.5)--
 (7.5,2.5)--
 (7.5,4.5)--
 (8.5,4.5)--
 (8.5,7.5)--
 (6.5,7.5)--
 (6.5,8.5)--
 (3.5,8.5)--
 (3.5,6.5)--
 (2.5,6.5)--
 (2.5,5.5)--
 (1.5,5.5)--
 (1.5,2.5);
  \draw[ultra thick](2,2)--(2,3);
  \draw[ultra thick](2,2)--(3,2);
  \draw[ultra thick](3,1)--(3,2);
  \draw[ultra thick](4,1)--(4,2);
  \draw[ultra thick](5,1)--(5,2);
  \draw[ultra thick](5,2)--(6,2);
  \draw[ultra thick](6,2)--(6,3);
  \draw[ultra thick](7,2)--(7,3);
  \draw[ultra thick](7,3)--(8,3);
  \draw[ultra thick](7,4)--(8,4);
  \draw[ultra thick](8,4)--(8,5);
  \draw[ultra thick](8,5)--(9,5);
  \draw[ultra thick](8,6)--(9,6);
  \draw[ultra thick](8,7)--(9,7);
  \draw[ultra thick](8,7)--(8,8);
  \draw[ultra thick](7,7)--(7,8);
  \draw[ultra thick](6,8)--(7,8);
  \draw[ultra thick](6,8)--(6,9);
  \draw[ultra thick](5,8)--(5,9);
  \draw[ultra thick](4,8)--(4,9);
  \draw[ultra thick](3,8)--(4,8);
  \draw[ultra thick](3,7)--(4,7);
  \draw[ultra thick](3,6)--(3,7);
  \draw[ultra thick](2,6)--(3,6);
  \draw[ultra thick](2,5)--(2,6);
  \draw[ultra thick](1,5)--(2,5);
  \draw[ultra thick](1,4)--(2,4);
  \draw[ultra thick](1,3)--(2,3);
  \node[blue] at (2.5,8.5) {$\widetilde{\mathcal{M}}_3$};
  \node[red] at (5,5) {$\mathcal{V}_4$};
\end{tikzpicture}
\caption{A 2D slice of an example of the 3D defect~$\widetilde{\mathcal{M}}_3$
(the broken line) and the interior of the defect, $\mathcal{V}_4$ (the shaded
area). The external gauge field~$U(x,\mu)$ on the links indicated by thick
ticks acquire nontrivial phases~$e^{\pm i\alpha/2}$ according to the rule
in~Eq.~\eqref{eq:(4.1)}.}
\label{fig:1}
\end{figure}

We first construct a defect which corresponds to the axial $U(1)$ (anomalous)
symmetry on the lattice. Since this is a 0-form symmetry acting on fermion
fields, we consider a co-dimension~1, i.e.\ 3D closed
surface~$\widetilde{\mathcal{M}}_3$ on the dual lattice of~$\Gamma$
(see~Fig.~\ref{fig:1}). In this paper, we assume that
$\widetilde{\mathcal{M}}_3$ is homologically trivial, i.e.\ there exists a 4D
region~$\mathcal{V}_4$ on~$\Gamma$ such that $\widetilde{\mathcal{M}}_3$ is
given by the boundary of~$\mathcal{V}_4$,
$\widetilde{\mathcal{M}}_3=\partial(\mathcal{V}_4)$. We also assume that
$\widetilde{\mathcal{M}}_3$ is endowed with a natural orientation from inside
to outside.

For a given defect~$\widetilde{\mathcal{M}}_3$, we then associate a particular
configuration of the external $U(1)$ lattice gauge field~$U(x,\mu)$ according
to the rule,
\begin{align}
   &U(x,\mu)
\notag\\
   &=
   \begin{cases}
   e^{+i\alpha/2}&
   \text{if the link $\langle x\to x+\Hat{\mu}\rangle$ intersects
   $\widetilde{\mathcal{M}}_3$ in positive direction},
   \\
   e^{-i\alpha/2}&
   \text{if the link $\langle x\to x+\Hat{\mu}\rangle$ intersects
   $\widetilde{\mathcal{M}}_3$ in negative direction},
   \\
   1&
   \text{otherwise},
   \end{cases}
\label{eq:(4.1)}
\end{align}
where $\alpha\in[0,2\pi)$ is a fixed real number. One can see that, for any
$\widetilde{\mathcal{M}}_3$ with properties assumed above,
$U(x,\mu)$~\eqref{eq:(4.1)} defines a pure-gauge $U(1)$ gauge field
as per~Eq.~\eqref{eq:(2.4)}.\footnote{From~Eq.~\eqref{eq:(4.1)}, it is obvious
that $U(x,\mu)$ possesses the vanishing field strength and all Wilson loops are
unity. Then, we may define $\Lambda(x)$ as the product of $U(y,\nu)$ along a
path from a certain fixed point~$o$ to~$x$. $\Lambda(x)$ is independent of the
chosen path and periodic on~$\Gamma$. Since
$\Lambda(x)U(x,\mu)\Lambda(x+\Hat{\mu})^{-1}=1$, $\Lambda(x)$ gives the gauge
transformation in~Eq.~\eqref{eq:(2.4)}.} This is expected because
Eq.~\eqref{eq:(4.1)} defines a lattice analogue of the Poincar\'e dual of a
homologically trivial surface~$\widetilde{\mathcal{M}}_3$.

\begin{figure}[htbp]
\centering
\begin{tikzpicture}[scale=0.9]
  \draw (-0.5,-0.5) grid[step=2] (4.5,4.5);
  \fill (2,2) circle(4pt);
  \draw[dashed,blue,ultra thick] (1,1)--(3,1)--(3,3)--(1,3)--(1,1);
  \draw[ultra thick] (2,2)--(4,2) node[below] {$e^{+i\alpha/2}$};
  \draw[ultra thick] (2,2)--(2,4) node[below left] {$e^{+i\alpha/2}$};
  \draw[ultra thick] (2,2)--(0,2) node[below right] {$e^{-i\alpha/2}$};
  \draw[ultra thick] (2,2)--(2,0) node[above right] {$e^{-i\alpha/2}$};
  \node[below right] at (2,2) {$y$};
  \node[blue] at (3.5,3.5) {$\widetilde{\mathcal{M}}_3$};
\end{tikzpicture}
\caption{A 2D slice of a defect~$\widetilde{\mathcal{M}}_3$ (the
broken line) surrounding a single site~$y$. The phases $e^{\pm\alpha/2}$ on
links are the values of~$U(x,\mu)$ assigned according to the
rule in~Eq.~\eqref{eq:(4.1)}.}
\label{fig:2}
\end{figure}
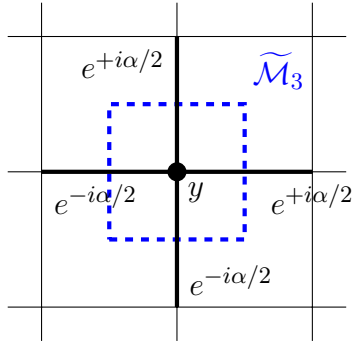

Now, we want to consider a deformation of the defect. We see that the
deformation can be realized by a gauge transformation on the external gauge
field~$U(x,\mu)$. Suppose $\widetilde{\mathcal{M}}_3$ is a defect that
surrounds a particular single site~$y\in\Gamma$ as per~Fig.~\ref{fig:2}; we
take the orientation of~$\widetilde{\mathcal{M}}_3$ such that the site~$y$ is
inside~$\widetilde{\mathcal{M}}_3$.
Then, it is easy to see that the $U(1)'$ gauge transformation
in~Eq.~\eqref{eq:(3.7)} with
\begin{equation}
   \Lambda(x)=e^{-i\alpha/2\delta_{xy}},
\label{eq:(4.2)}
\end{equation}
eliminates the defect, i.e.\ all $U(x,\mu)$ becomes unity. From this
observation, it is clear that one can freely deform a general
defect~$\widetilde{\mathcal{M}}_3$ by repeatedly applying this kind of gauge
transformations site by site.

The functional integral over the fermion fields that we have defined in the
previous section is, however, not invariant under the $U(1)'$ gauge
transformation~\eqref{eq:(4.2)}; we must have an anomaly. The finite gauge
transformation~\eqref{eq:(4.2)} can be constructed by accumulating
infinitesimal gauge transformations parametrized by (see Eq.~\eqref{eq:(3.26)})
\begin{equation}
   \eta_\mu'(x)=-\partial_\mu\Omega(x),\qquad
   \Omega(x)=-\frac{1}{2}\delta\alpha\delta_{xy}.
\label{eq:(4.3)}
\end{equation}
Then, Eq.~\eqref{eq:(3.31)} yields
\begin{equation}
   \delta_\eta\langle\mathcal{O}\rangle_{\mathrm{F}}
   =\langle\delta_\eta\mathcal{O}\rangle_{\mathrm{F}}
   +i\delta\alpha\gamma
   \epsilon_{\mu\nu\rho\sigma}
   f_{\mu\nu}(y)f_{\rho\sigma}(y+\Hat{\mu}+\Hat{\nu})
   \langle\mathcal{O}\rangle_{\mathrm{F}},
\label{eq:(4.4)}
\end{equation}
where, for the fermion fields inside the expectation value (see
Eq.~\eqref{eq:(3.33)}),
\begin{equation}
   \delta_\eta\psi(x)=\frac{i}{2}\delta\alpha\psi(x)\delta_{xy},\qquad
   \delta_\eta\Bar{\psi}(x)
   =-\frac{i}{2}\delta\alpha\Bar{\psi}(x)\delta_{xy},
\label{eq:(4.5)}
\end{equation}
where we have noted $T_{\alpha\beta}=-\delta_{\alpha\beta}$. Integrating
Eq.~\eqref{eq:(4.4)} with respect to~$\alpha$, we thus have
\begin{equation}
   \langle\mathcal{O}\rangle_{\mathrm{F}}^{\widetilde{\mathcal{M}}_3}
   =\exp
   \left[
   -i\alpha\gamma
   \epsilon_{\mu\nu\rho\sigma}
   f_{\mu\nu}(y)f_{\rho\sigma}(y+\Hat{\mu}+\Hat{\nu})
   \right]
   \langle\mathcal{O}^\alpha\rangle_{\mathrm{F}},
\label{eq:(4.6)}
\end{equation}
where, on the right-hand side, fermion fields within the
operator~$\mathcal{O}^\alpha$ are given by\footnote{Since we introduced $\psi$
as a left-handed Weyl fermion, in terms of the Dirac fermion, these
transformations correspond to $\psi(x)^\alpha=e^{i(\alpha/2)\gamma_5}\psi(x)$
and~$\Bar{\psi}(x)^\alpha=\Bar{\psi}(x)e^{i(\alpha/2)\gamma_5}$. The reason of the
factor~$1/2$ in the rotation angle is that the axial rotation
with~$\alpha=2\pi$, $\psi(x)^{2\pi}=-\psi(x)$
and~$\Bar{\psi}(x)^{2\pi}=-\Bar{\psi}(x)$, corresponds to a vectorial gauge
transformation and thus is regarded as the identity transformation.}
\begin{equation}
   \psi(x)^\alpha=e^{-i\alpha/2\delta_{xy}}\psi(x),\qquad
   \Bar{\psi}(x)^\alpha=\Bar{\psi}(x)e^{i\alpha/2\delta_{xy}}.
\label{eq:(4.7)}
\end{equation}
In Eq.~\eqref{eq:(4.6)}, the left-hand side is the expectation value with the
defect~$\widetilde{\mathcal{M}}_3$ surrounding a single site~$y$
(Fig.~\ref{fig:2}). The right-hand side is the expectation value without the
defect. By repeatedly using the relation~\eqref{eq:(4.6)} site by site,
therefore, for an arbitrary
defect~$\widetilde{\mathcal{M}}_3=\partial\mathcal{V}_4$, we have
\begin{equation}
   \langle\mathcal{O}\rangle_{\mathrm{F}}^{\widetilde{\mathcal{M}}_3}
   =\exp
   \left[
   -i\alpha\gamma\sum_{x\in\mathcal{V}_4}
   \epsilon_{\mu\nu\rho\sigma}
   f_{\mu\nu}(x)f_{\rho\sigma}(x+\Hat{\mu}+\Hat{\nu})
   \right]
   \langle\mathcal{O}^\alpha\rangle_{\mathrm{F}},
\label{eq:(4.8)}
\end{equation}
where, on the right-hand side,
\begin{equation}
   \psi(x)^\alpha
   =\begin{cases}
   e^{-i\alpha/2}\psi(x)&\text{for $x\in\mathcal{V}_4$},\\ 
   \psi(x)&\text{otherwise}.\\ 
   \end{cases}\qquad
   \Bar{\psi}(x)^\alpha
   =\begin{cases}
   \Bar{\psi}(x)e^{i\alpha/2}&\text{for $x\in\mathcal{V}_4$},\\ 
   \Bar{\psi}(x)&\text{otherwise}.\\ 
   \end{cases}
\label{eq:(4.9)}
\end{equation}
Equation~\eqref{eq:(4.8)} is nothing but the anomalous chiral Ward--Takahashi
identity. That is, we may deform away the defect~$\widetilde{\mathcal{M}}_3$.
This induces the chiral rotation of the angle~$\alpha/2$ on fermion fields on
the sites swept by the deformation. The deformation, however, also leaves the
effect of the axial anomaly. That is, the defect is not quite topological for
this anomalous symmetry. We may rewrite Eq.~\eqref{eq:(4.8)} as
\begin{equation}
   \left\langle U_\alpha(\widetilde{\mathcal{M}}_3)\mathcal{O}
   \right\rangle_{\mathrm{F}}
   :=\langle\mathcal{O}\rangle_{\mathrm{F}}^{\widetilde{\mathcal{M}}_3}
   \exp
   \left[
   i\alpha\gamma\sum_{x\in\mathcal{V}_4}
   \epsilon_{\mu\nu\rho\sigma}
   f_{\mu\nu}(x)f_{\rho\sigma}(x+\Hat{\mu}+\Hat{\nu})
   \right]
   =\langle\mathcal{O}^\alpha\rangle_{\mathrm{F}}
\label{eq:(4.10)}
\end{equation}
and define the symmetry operator~$U_\alpha(\widetilde{\mathcal{M}}_3)$ in terms
of the functional integral. $U_\alpha(\widetilde{\mathcal{M}}_3)$ is topological
because the most right-hand side of~Eq.~\eqref{eq:(4.10)} is independent of the
precise form of the defect~$\widetilde{\mathcal{M}}_3$. We note that the
combination in~Eq.~\eqref{eq:(4.10)} is invariant under the physical $U(1)$
lattice gauge transformations; the physical $U(1)$ does not have the anomaly
as~Eq.~\eqref{eq:(3.31)} shows and the field strength $f_{\mu\nu}(x)$ is also
invariant under~$U(1)$.

\begin{figure}[htbp]
\centering
\begin{tikzpicture}[scale=0.7]
  \fill[red!20]
 (2,3)--
 (3,3)--
 (3,2)--
 (4,2)--
 (4,2)--
 (5,2)--
 (5,3)--
 (7,3)--
 (7,5)--
 (8,5)--
 (8,7)--
 (6,7)--
 (6,8)--
 (4,8)--
 (4,6)--
 (3,6)--
 (3,5)--
 (2,5)--
 (2,3);
  \draw (-0.5,-0.5) grid[step=1] (10.5,10.5);
  \draw[ultra thick]
 (2,3)--
 (3,3)--
 (3,2)--
 (4,2)--
 (4,2)--
 (6,2)--
 (6,3)--
 (8,3)--
 (8,5)--
 (9,5)--
 (9,8)--
 (7,8)--
 (7,9)--
 (4,9)--
 (4,7)--
 (3,7)--
 (3,6)--
 (2,6)--
 (2,3);
  \draw[dashed,blue,ultra thick]
 (1.5,2.5)--
 (2.5,2.5)--
 (2.5,1.5)--
 (3.5,1.5)--
 (3.5,1.5)--
 (5.5,1.5)--
 (5.5,2.5)--
 (7.5,2.5)--
 (7.5,4.5)--
 (8.5,4.5)--
 (8.5,7.5)--
 (6.5,7.5)--
 (6.5,8.5)--
 (3.5,8.5)--
 (3.5,6.5)--
 (2.5,6.5)--
 (2.5,5.5)--
 (1.5,5.5)--
 (1.5,2.5);
  \node[blue] at (2.5,8.5) {$\widetilde{\mathcal{M}}_3$};
  \node at (7.5,9.5) {$\mathcal{M}_3$};
  \node[red] at (5,5) {$\mathcal{V}_4$};
\end{tikzpicture}
\caption{A 2D slice of the defect~$\widetilde{\mathcal{M}}_3$ (the
broken line) and the interior of the defect, $\mathcal{V}_4$ (the shaded area).
The corresponding 3-surface~$\mathcal{M}_3$ (the bold line) defines the
boundary variables.}
\label{fig:3}
\end{figure}
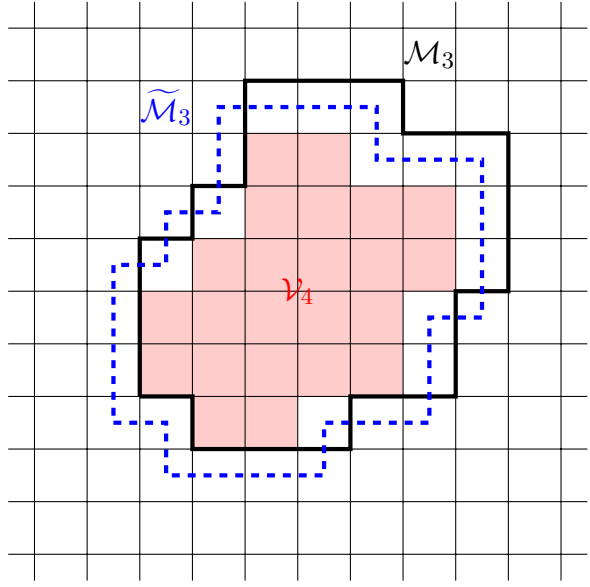

We now note from~Eq.~\eqref{eq:(2.13)} that the topological density appearing
in~Eq.~\eqref{eq:(4.10)} can be written as\footnote{A quick way to find this
kind of expression is to employ the non-commutative differential
calculus~\cite{Fujiwara:1999fi}.}
\begin{align}
   &\epsilon_{\mu\nu\rho\sigma}
   f_{\mu\nu}(x)f_{\rho\sigma}(x+\Hat{\mu}+\Hat{\nu})
\notag\\
   &=2\epsilon_{\mu\nu\rho\sigma}\partial_\mu
   \left[a_\nu(x)f_{\rho\sigma}(x+\Hat{\nu})
   +2\pi z_{\nu\rho}(x)a_\sigma(x+\Hat{\nu}+\Hat{\rho})\right]
\notag\\
   &\qquad{}
   +4\pi^2\epsilon_{\mu\nu\rho\sigma}
   z_{\mu\nu}(x)z_{\rho\sigma}(x+\Hat{\mu}+\Hat{\nu}).
\label{eq:(4.11)}
\end{align}
The first term on the right-hand side is a total difference on the lattice
and the combination in the square brackets would be regarded as a lattice
counterpart of the Chern--Simons 3-form. In fact, since $z_{\mu\nu}(x)$ in the
last term are integers, the change of~Eq.~\eqref{eq:(4.11)} under an
infinitesimal variation of the lattice gauge field~$u(x,\mu)$ depends only on
the variation of~$u(x,\mu)$ on a 3D closed surface corresponding
to~$\widetilde{\mathcal{M}}_3$; we denote this closed 3-surface on the original
lattice as~$\mathcal{M}_3$. See Fig.~\ref{fig:3}.
In what follows, we call link variables~$u(x,\mu)$ with the
link~$\langle x\to x+\Hat{\mu}\rangle\in\mathcal{M}_3$ boundary variables,
while $u(x,\mu)$
with~$\langle x\to x+\Hat{\mu}\rangle\in\interior(\mathcal{M}_3)$ bulk
variables. Substituting Eq.~\eqref{eq:(4.11)} into~Eq.~\eqref{eq:(4.10)},
the first term on the right-hand side of~Eq.~\eqref{eq:(4.11)} gives rise to
$\sum_{x\in\mathcal{M}_3}\epsilon_{\mu\nu\rho\sigma}
[a_\nu(x)f_{\rho\sigma}(x+\Hat{\nu})+2\pi z_{\nu\rho}(x)a_\sigma(x+\Hat{\nu}+\Hat{\rho})]$, a quantity that depends on the boundary variables only. We propose to
regard this combination as the $U(1)$ Chern--Simons term on the lattice; in
other words, we regard the quantity in the square brackets
of~Eq.~\eqref{eq:(4.10)} as a possible definition of the $U(1)$ Chern--Simons
term associated with the
defect~$\widetilde{\mathcal{M}}_3=\partial(\mathcal{V}_4)$.

A quite important characterization of the Chern--Simons action is its gauge
\emph{non}invariance. The Chern--Simons action on a closed manifold can be
regraded gauge invariant only when the level is an integer. With this fact and
the above idea on the lattice Chern--Simons term in mind, in what follows, we
consider the invariance of~Eq.~\eqref{eq:(4.10)} under the gauge transformation
on \emph{the boundary variables only}. As we will analyze below,
Eq.~\eqref{eq:(4.10)} is not invariant such a transformation. We regard this
noninvariance as a good indication because the Chern--Simons action in
continuum is not gauge invariant in general. How to make the expression
invariant also under the gauge transformation on the boundary variables only is
the subject of the next subsection.

\subsection{Gauge average and the projection operator for magnetic fluxes on
the defect}
\label{sec:4.2}
The symmetry operator~$U_\alpha(\widetilde{\mathcal{M}}_3)$
in~Eq.~\eqref{eq:(4.10)} is topological. However, since Eq.~\eqref{eq:(4.10)}
is not invariant under gauge transformations on \emph{the boundary variables
only\/} as we will see below, $U_\alpha(\widetilde{\mathcal{M}}_3)$, when seeing
as an object depending only on the boundary variables, is not gauge invariant.
In this sense, $U_\alpha(\widetilde{\mathcal{M}}_3)$ may be regarded as an
unphysical operator.

To make the symmetry operator also invariant under gauge transformations on the
boundary variables only, we imitate the prescription by
Karasik~\cite{Karasik:2022kkq} in continuum and integrate
$U_\alpha(\widetilde{\mathcal{M}}_3)$ over gauge transformations on the boundary
variables only.\footnote{The idea of the gauge average itself is quite general;
see, e.g., Refs.~\cite{Forster:1980dg,Harada:1986wb}.} That is, we
consider the average over gauge transformations:
\begin{align}
   &\left\langle\Tilde{U}_\alpha(\widetilde{\mathcal{M}}_3)\mathcal{O}
   \right\rangle_{\mathrm{F}}
\notag\\
   &:=
   \langle\mathcal{O}\rangle_{\mathrm{F}}^{\widetilde{\mathcal{M}}_3}
   \int\mathrm{D}[\lambda]\,
   \exp
   \left[
   i\alpha\gamma\sum_{x\in\mathcal{V}_4}
   \epsilon_{\mu\nu\rho\sigma}
   f_{\mu\nu}(x)f_{\rho\sigma}(x+\Hat{\mu}+\Hat{\nu})
   \right]^\lambda
   =\langle\mathcal{O}^\alpha\rangle_{\mathrm{F}},
\label{eq:(4.12)}
\end{align}
where~$\mathrm{D}[\lambda]$ is the integration over gauge transformations on
the boundary variables only; the precise definition of the measure is given
in~Appendix~\ref{sec:B}.\footnote{Unfortunately, in order to reproduce the
prescription of~Ref.~\cite{Karasik:2022kkq}, we have to assume an elaborated
form of the integration measure~$\int\mathrm{D}[\lambda]$ because
the integration volume can depend on the winding numbers; this point was
overlooked in the first version of the present paper and we would like to thank
Yuya Tanizaki for pointing this out to us.} The symbol~$[\phantom{X}]^\lambda$
indicates the gauge transformation on the boundary variables only corresponding
to the gauge average of the Chern--Simons action on the defect in the
continuum~\cite{Karasik:2022kkq}.

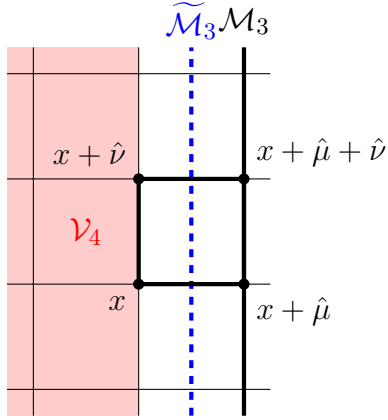
\begin{figure}[htbp]
\centering
\begin{tikzpicture}[scale=0.7]
  \fill[red!20] (-0.5,-0.5)--(2,-0.5)--(2,6.5)--(-0.5,6.5);
  \draw (-0.5,-0.5) grid[step=2] (4.5,6.5);
  \fill (2,2) circle(3pt);
  \fill (4,2) circle(3pt);
  \fill (4,4) circle(3pt);
  \fill (2,4) circle(3pt);
  \draw[dashed,blue,ultra thick] (3,-0.5)--(3,6.5);
  \draw[ultra thick] (4,-0.5)--(4,6.5);
  \draw[ultra thick] (2,2)--(4,2) node[below right] {$x+\Hat{\mu}$};
  \draw[ultra thick] (4,2)--(4,4) node[above right] {$x+\Hat{\mu}+\Hat{\nu}$};
  \draw[ultra thick] (4,4)--(2,4) node[above left] {$x+\Hat{\nu}$};
  \draw[ultra thick] (2,4)--(2,2) node[below left] {$x$};
  \node[blue] at (3,7) {$\widetilde{\mathcal{M}}_3$};
  \node[red] at (1,3) {$\mathcal{V}_4$};
  \node at (4,7) {$\mathcal{M}_3$};
\end{tikzpicture}
\caption{A plaquette extending between $\mathcal{M}_3$ and~$\mathcal{V}_4$.
Only the gauge field~$u(x+\Hat{\mu},\nu)$ on the
link~$\langle x+\Hat{\mu}\to x+\Hat{\mu}+\Hat{\nu}\rangle$ is the boundary
variable which is subject to the gauge transformation in~Eq.~\eqref{eq:(4.18)}.
The plaquette with the replacement $x\to x+\Hat{\rho}+\Hat{\sigma}$ also
contributes to~Eq.~\eqref{eq:(4.22)}.}
\label{fig:4}
\end{figure}

Suppose that, in the plaquette,
\begin{equation}
   u(x,\mu)u(x+\Hat{\mu},\nu)u(x+\Hat{\nu},\mu)^{-1}u(x,\nu)^{-1},
\label{eq:(4.13)}
\end{equation}
only $u(x+\Hat{\mu},\nu)$ is the boundary variable, while others are bulk
variables. See~Fig.~\ref{fig:4}.
Then, we consider the gauge transformation acting only on the boundary
variable,
\begin{equation}
   u(x+\Hat{\mu},\nu)\to
   \lambda(x+\Hat{\mu})
   u(x+\Hat{\mu},\nu)
   \lambda(x+\Hat{\mu}+\Hat{\nu})^{-1}
   =e^{-i\phi(x+\Hat{\mu})}u(x+\Hat{\mu},\nu)e^{i\phi(x+\Hat{\mu}+\Hat{\nu})},
\label{eq:(4.14)}
\end{equation}
where we have set
\begin{equation}
   \lambda(x)=e^{-i\phi(x)},\qquad-\pi<\phi(x)\leq\pi.
\label{eq:(4.15)}
\end{equation}
To realize the picture of~Ref.~\cite{Karasik:2022kkq}, however, it turns out
that we have to impose a certain smoothness condition on possible gauge
transformations~$\lambda$ to be integrated in~Eq.~\eqref{eq:(4.12)}. That is,
setting,
\begin{equation}
   -\pi<\frac{1}{i}\ln\left[
   e^{-i\phi(x+\Hat{\mu})}e^{i\phi(x+\Hat{\mu}+\Hat{\nu})}\right]
   =\partial_\nu\phi(x+\Hat{\mu})+2\pi l_\nu(x+\Hat{\mu})\leq\pi,
\label{eq:(4.16)}
\end{equation}
where $l_\nu(x+\Hat{\mu})\in\mathbb{Z}$, we require that the gauge
transformation function~$\phi(x)$ varies sufficiently smoothly on the lattice,
satisfying
\begin{equation}
   \sup_{x,\mu}\left|
   \partial_\mu\phi(x)+2\pi l_\mu(x)
   \right|<\epsilon',\qquad
   0<\epsilon'<\frac{\pi}{6}.
\label{eq:(4.17)}
\end{equation}
The transformation~\eqref{eq:(4.14)} then induces the change of the field
strength, 
\begin{align}
   f_{\mu\nu}(x)
   &\to
   \frac{1}{i}
   \ln
   \left[u(x,\mu)
   e^{-i\phi(x+\Hat{\mu})}
   u(x+\Hat{\mu},\nu)
   e^{i\phi(x+\Hat{\mu}+\Hat{\nu})}
   u(x+\Hat{\nu},\mu)^{-1}u(x,\nu)^{-1}
   \right]
\notag\\
   &=f_{\mu\nu}(x)+\partial_\nu\phi(x+\Hat{\mu})
   +2\pi l_\nu(x+\Hat{\mu}).
\label{eq:(4.18)}
\end{align}
Note that the right-hand side is within the desired branch
$-\pi<f_{\mu\nu}(x)+\partial_\nu\phi(x+\Hat{\mu})
+2\pi l_\nu(x+\Hat{\mu})\leq\pi$ because of~Eqs.~\eqref{eq:(2.10)}
and~\eqref{eq:(4.17)}.\footnote{Although in~Eq.~\eqref{eq:(4.18)} we have
considered the case in which only one link variable of a plaquette is belonging
to the boundary variables, there exist cases in which two, three, and four link
variables of a plaquette are belonging to the boundary variables. The
bound $\epsilon'<\pi/6$ is designed to include these possibilities.} Since
Eq.~\eqref{eq:(4.17)} implies
\begin{equation}
   \left|
   \epsilon_{\mu\nu\rho\sigma}
   \partial_\rho l_\sigma(x)
   \right|<\frac{4}{2\pi}\epsilon'
   <1,
\label{eq:(4.19)}
\end{equation}
and $l_\mu(x)$ are integers, the integer field~$l_\mu(x)$ is flat (i.e.,
rotation-free):
\begin{equation}
   \epsilon_{\mu\nu\rho\sigma}
   \partial_\rho l_\sigma(x)=0.
\label{eq:(4.20)}
\end{equation}

The motivation for the gauge average in~Eq.~\eqref{eq:(4.12)} is to make the
lattice analogue of the Chern--Simons term invariant under gauge
transformations of boundary variables only. The invariance is archived,
however, in a restricted sense. Under the gauge transformation~$\lambda'$ on
boundary variables only,
\begin{align}
   &\left\langle\Tilde{U}_\alpha(\widetilde{\mathcal{M}}_3)\mathcal{O}
   \right\rangle_{\mathrm{F}}^{\lambda'}
\notag\\
   &=
   \langle\mathcal{O}\rangle_{\mathrm{F}}^{\widetilde{\mathcal{M}}_3}
   \int\mathrm{D}[\lambda]\,
   \exp
   \left[
   i\alpha\gamma\sum_{x\in\mathcal{V}_4}
   \epsilon_{\mu\nu\rho\sigma}
   f_{\mu\nu}(x)f_{\rho\sigma}(x+\Hat{\mu}+\Hat{\nu})
   \right]^{\lambda\lambda'}.
\label{eq:(4.21)}
\end{align}
From this, one might think that $\lambda'$ can be absorbed by the
redefinition~$\lambda\to\lambda(\lambda')^{-1}$ and the symmetry operator is
made invariant. However, since our consideration below is based on the
flatness~\eqref{eq:(4.20)}, the change of the gauge transformation
$\lambda\to\lambda\lambda'$ in~Eq.~\eqref{eq:(4.21)} should not influence the
flatness~\eqref{eq:(4.20)}. Also, our measure $\int\mathrm{D}[\lambda]$
in~Appendix~\ref{sec:B} is not invariant under such a shift,
$\lambda\to\lambda(\lambda')^{-1}$. It is clear, however that if $\lambda'$ is
sufficiently smooth, $|\partial_\mu\phi'(x)|\ll1$
for~$\lambda'(x)=e^{-i\phi'(x)}$, then the integer field~$l_\mu(x)$ defined
in~Eq.~\eqref{eq:(4.16)} does not change under~$\lambda\to\lambda\lambda'$. We
thus require this sufficient smoothness for~$\lambda'$
in~Eq.~\eqref{eq:(4.21)}. The gauge invariance of our lattice Chern--Simons
term on the defect as a function of boundary variables is restricted in this
sense. We expect that this point is not physically problematic because in
continuum limit only differentiable gauge transformations are relevant; in the
cutoff scale, these satisfy $|\partial_\mu\phi'(x)|\ll1$.\footnote{Since if we
restore the lattice spacing, this reads $|\partial_\mu\phi'(x)|\ll1/a$.}

Let us now fix a direction~$\mu$ and assume that $\widetilde{\mathcal{M}}_3$
is perpendicular to~$\Hat{\mu}$ and intersects a
link~$\langle x\to x+\Hat{\mu}\rangle$ as~Fig.~\ref{fig:4}. In what follows,
when necessary, we consider the case~$\widetilde{\mathcal{M}}_3=T^3$ as an
explicit example, although we think that general~$\widetilde{\mathcal{M}}_3$
can be treated by appropriate modifications of expressions. Then, under the
gauge transformation on the boundary variables, the topological density
$\epsilon_{\mu\nu\rho\sigma}f_{\mu\nu}(x)f_{\rho\sigma}(x+\Hat{\mu}+\Hat{\nu})$~\eqref{eq:(4.11)} with~$x+\Hat{\mu}\in\mathcal{M}_3=T^3$ changes by (here $\mu$ is
fixed and not summed over; the 3-surface~$\mathcal{M}_3$ is extending in the
directions of $\nu$, $\rho$, and~$\sigma$)
\begin{align}
   &2\epsilon_{\mu\nu\rho\sigma}
   \left[
   \partial_\nu\phi(x+\Hat{\mu})+2\pi l_\nu(x+\Hat{\mu})
   \right]f_{\rho\sigma}(x+\Hat{\mu}+\Hat{\nu})
\notag\\
   &\qquad{}
   +2\epsilon_{\rho\sigma\mu\nu}f_{\rho\sigma}(x)
   \left[
   \partial_\nu\phi(x+\Hat{\mu}+\Hat{\rho}+\Hat{\sigma})
   +2\pi l_\nu(x+\Hat{\mu}+\Hat{\rho}+\Hat{\sigma})
   \right]
\notag\\
   &=4\pi\epsilon_{\mu\nu\rho\sigma}
   \left[
   l_\nu(x+\Hat{\mu})f_{\rho\sigma}(x+\Hat{\mu}+\Hat{\nu})
   +f_{\rho\sigma}(x)l_\nu(x+\Hat{\mu}+\Hat{\rho}+\Hat{\sigma})
   \right]
\notag\\
   &\qquad{}
   +2\epsilon_{\mu\nu\rho\sigma}
   \partial_\nu\left[
   \phi(x+\Hat{\mu})f_{\rho\sigma}(x+\Hat{\mu})
   +f_{\rho\sigma}(x)\phi(x+\Hat{\mu}+\Hat{\rho}+\Hat{\sigma})
   \right],
\label{eq:(4.22)}
\end{align}
where we have used the Bianchi identity~\eqref{eq:(2.16)}. In deriving this,
we have noted that $f_{\alpha\beta}(x)$ is gauge invariant if $\alpha$
and~$\beta$ do not contain~$\mu$ and $f_{\mu\alpha}(x)$
with~$x+\Hat{\mu}\in\mathcal{M}_3$ receives a nontrivial gauge transformation.
Then, under the sum over $x+\Hat{\mu}\in\mathcal{M}_3$, the last line vanishes.

At this point, we note that the flatness~\eqref{eq:(4.20)} allows us to
introduce a ``scalar potential'' $\varphi(x)$ for the integer vector
field~$l_\nu(x)$ as
(here, sites $o$, $x$ and~$y$ all are belonging to~$\mathcal{M}_3$),
\begin{equation}
   l_\nu(x)=\partial_\nu\varphi(x),\qquad
   \varphi(x)
   =\sum_{\langle y\to y+\rho\rangle
   \in\text{a path connecting $o$ and $x$ in~$\mathcal{M}_3$}}
   l_\rho(y),
\label{eq:(4.23)}
\end{equation}
where we have assumed a certain fixed point~$o\in\mathcal{M}_3$. $\varphi(x)$
is invariant under any deformation of the path because of the
flatness~\eqref{eq:(4.20)} and then the first relation obviously holds.
$\varphi(x)$ is, however, not single-valued in~$\mathcal{M}_3$ because of a
possibility that the path winds nontrivial cycles of~$\mathcal{M}_3$. When
$\mathcal{M}_3=T^3$ with the size~$\ell$, the path can wind one of three cycles
of~$T^3$. Under a single winding to the $\nu$~direction, $\varphi(x)$ acquires
an additional integer
\begin{equation}
   k_\nu=\sum_{s=0}^{\ell-1}l_\nu(x+s\Hat{\nu})
   =\sum_{s=0}^{\ell-1}\left[\frac{1}{2\pi}\partial_\nu\phi(x+s\Hat{\nu})
   +l_\nu(x+s\Hat{\nu})\right].
\label{eq:(4.24)}
\end{equation}
This number is \emph{independent} of the site~$x$ in~$\mathcal{M}_3$ again
because of the flatness~\eqref{eq:(4.20)}. From the
definition~\eqref{eq:(4.24)}, one sees that $k_\nu$ is the number of how many
times $e^{-i\phi(x)}e^{+i\phi(x+\Hat{\nu})}$ winds around~$U(1)$ as $x$ goes along
the $\nu$~direction. The smoothness~\eqref{eq:(4.19)} is required to make this
winding number well-defined on the lattice. Therefore, if we integrate over
possible smooth gauge transformations in~Eq.~\eqref{eq:(4.12)}, one has a
summation over integers~$k_\nu$ for each~$\nu$. Because
of~Eq.~\eqref{eq:(4.17)}, for finite~$\ell$, the winding number is bounded by
\begin{equation}
   \left|k_\nu\right|<\frac{\epsilon'}{2\pi}\ell.
\label{eq:(4.25)}
\end{equation}

Finally, using the representation~\eqref{eq:(4.23)} in~Eq.~\eqref{eq:(4.22)},
we have
\begin{align}
   &\sum_{x\in\mathcal{V}_4}
   \epsilon_{\mu\nu\rho\sigma}
   f_{\mu\nu}(x)f_{\rho\sigma}(x+\Hat{\mu}+\Hat{\nu})
\notag\\
   &\to\sum_{x\in\mathcal{V}_4}
   \epsilon_{\mu\nu\rho\sigma}
   f_{\mu\nu}(x)f_{\rho\sigma}(x+\Hat{\mu}+\Hat{\nu})
\notag\\
   &\qquad{}
   +4\pi\sum_{x+\Hat{\mu}\in\mathcal{M}_3}
   \epsilon_{\mu\nu\rho\sigma}
   \partial_\nu
   \left[\varphi(x+\Hat{\mu})f_{\rho\sigma}(x+\Hat{\mu})
   +f_{\rho\sigma}(x)\varphi(x+\Hat{\rho}+\Hat{\sigma})
   \right]
\notag\\
   &=\sum_{x\in\mathcal{V}_4}
   \epsilon_{\mu\nu\rho\sigma}
   f_{\mu\nu}(x)f_{\rho\sigma}(x+\Hat{\mu}+\Hat{\nu})
\notag\\
   &\qquad{}
   +4\pi k_\nu\sum_{x+\Hat{\mu}\in\mathcal{M}_3,x_\nu=0}
   \epsilon_{\mu\nu\rho\sigma}
   \left[f_{\rho\sigma}(x+\Hat{\mu})+f_{\rho\sigma}(x)\right],
\label{eq:(4.26)}
\end{align}
where we have used the Bianchi identity~\eqref{eq:(2.16)} and the fact that
$\varphi(x)$ acquires the number~$k_\nu$ after a single return around the
$\nu$~direction. From this, we see that the integration over smooth gauge
transformations in~Eq.~\eqref{eq:(4.12)} produces a factor,
\begin{align}
   &\int\mathrm{D}[\lambda]\,
   \exp\left\{
   4\pi i\alpha\gamma
   k_\nu
   \sum_{x+\Hat{\mu}\in\mathcal{M}_3,x_\nu=0}\epsilon_{\mu\nu\rho\sigma}
   \left[
   f_{\rho\sigma}(x+\Hat{\mu})+f_{\rho\sigma}(x)
   \right]
   \right\}
\notag\\
   &=\int\mathrm{D}[\lambda]\,
   \exp\left[
   8\pi i\alpha\gamma
   k_\nu
   \sum_{x\in\mathcal{M}_2^\nu}
   \epsilon_{\mu\nu\rho\sigma}f_{\rho\sigma}(x)
   \right],
\label{eq:(4.27)}
\end{align}
where $\mathcal{M}_2^\nu$ is a closed surface such that
$\mathcal{M}_3=S^1\times\mathcal{M}_2^\nu$ with $S^1$ being the direction
of~$\nu$. The precise form of~$\mathcal{M}_2^\nu$ does not matter because of
the Bianchi identity~\eqref{eq:(2.16)}. That is, a directional sum
of~$f_{\mu\nu}(x)$ over a surface of a 3D cube vanishes and one can freely
deform $\mathcal{M}_2^\nu$ as far as it can be done by the adding/subtracting
of a cube. As noted above, with the integration measure
in~Appendix~\ref{sec:B}, the integration over smooth gauge transformations
becomes the sum over the winding number~$k_\nu$. Introducing the
function~$\delta_\ell(x)$ by
\begin{equation}
   \sum_{k=-[\epsilon'\ell/(2\pi)]}^{[\epsilon'\ell/(2\pi)]}
   e^{ikx}=2\pi\sum_{n=-\infty}^\infty\delta_\ell(x-2\pi n),
\label{eq:(4.28)}
\end{equation}
we know
\begin{equation}
   \delta_\ell(x)\stackrel{\ell\to\infty}{\to}\delta(x).
\label{eq:(4.29)}
\end{equation}
Therefore, the gauge average gives rise to, for each~$\nu$,
\begin{equation}
   \delta_\ell\left(
   \frac{\alpha}{2\pi}\frac{1}{4\pi}
   \sum_{x\in\mathcal{M}_2^\nu}
   \epsilon_{\mu\nu\rho\sigma}f_{\rho\sigma}(x)
   -\mathbb{Z}
   \right),
\label{eq:(4.30)}
\end{equation}
since $\gamma=-1/(32\pi^2)$ (recall~Eq.~\eqref{eq:(3.32)}). This implies that,
when $\alpha/(2\pi)$ is an irrational number,
\begin{equation}
   \frac{1}{4\pi}
   \sum_{x\in\mathcal{M}_2^\nu}
   \epsilon_{\mu\nu\rho\sigma}f_{\rho\sigma}(x)
   \stackrel{\ell\to\infty}{\to}0,
\label{eq:(4.31)}
\end{equation}
i.e., any magnetic fluxes along~$\mathcal{M}_2^\nu$ are not allowed, while,
when $\alpha/(2\pi)$ is a rational number, $p/N$, where $p$ and~$N$ are
co-prime integers,
\begin{equation}
   \frac{1}{4\pi}
   \sum_{x\in\mathcal{M}_2^\nu}
   \epsilon_{\mu\nu\rho\sigma}
   f_{\rho\sigma}(x)
   \stackrel{\ell\to\infty}{\to}N\mathbb{Z}.
\label{eq:(4.32)}
\end{equation}
This precisely corresponds to the constraint on the magnetic flux in the
continuum obtained in~Ref.~\cite{Karasik:2022kkq}:
\begin{equation}
   \frac{1}{2\pi}\int_{\mathcal{M}_2}da=N\mathbb{Z}.
\label{eq:(4.33)}
\end{equation}
These observations show that, for $\ell\to\infty$, we can write the symmetry
operator defined by~Eq.~\eqref{eq:(4.12)} as
\begin{equation}
   \Tilde{U}_\alpha(\widetilde{\mathcal{M}}_3)
   =U_\alpha(\widetilde{\mathcal{M}}_3)P_\alpha(\widetilde{\mathcal{M}}_3),
\label{eq:(4.34)}
\end{equation}
where $P_\alpha(\widetilde{\mathcal{M}}_3)$ is a projection operator on the
subspace of allowed magnetic fluxes in~$\widetilde{\mathcal{M}}_3$. From this
expression, it is obvious that $\Tilde{U}_\alpha(\widetilde{\mathcal{M}}_3)$
represents a noninvertible symmetry because it contains a projection operator.
Gathering Eqs.~\eqref{eq:(2.5)}, \eqref{eq:(4.10)},
and~\eqref{eq:(4.30)}, in terms of the functional integral, the lattice
realization of the axial $U(1)$ noninvertible symmetry is thus given by
\begin{align}
   &\left\langle\Tilde{U}_{p/N}(\widetilde{\mathcal{M}}_3)\mathcal{O}
   \right\rangle
\notag\\
   &=\frac{1}{\mathcal{Z}}
   \int\mathrm{D}[u]\,e^{-S_\mathrm{G}}\,
   \exp
   \left[
   -\frac{ip}{16\pi N}\sum_{x\in\mathcal{V}_4}
   \epsilon_{\mu\nu\rho\sigma}
   f_{\mu\nu}(x)f_{\rho\sigma}(x+\Hat{\mu}+\Hat{\nu})
   \right]
\notag\\
   &\qquad{}
   \times\prod_{\nu\neq\mu}\sum_{n=-\infty}^\infty
   \delta_\ell\left(
   \frac{1}{4\pi}
   \sum_{x\in\mathcal{M}_2^\nu}
   \epsilon_{\mu\nu\rho\sigma}
   f_{\rho\sigma}(x)
   -Nn\right)
   \langle\mathcal{O}\rangle_{\mathrm{F}}^{\widetilde{\mathcal{M}}_3},
\label{eq:(4.35)}
\end{align}
where $\mu$ is the direction normal to~$\mathcal{M}_3=\partial(\mathcal{V}_4)$;
$\nu$ is the direction of~$S^1$ in the
decomposition~$\mathcal{M}_3=S^1\times\mathcal{M}_2^\nu$. The functional
integral over the fermion,
$\langle\mathcal{O}\rangle_{\mathrm{F}}^{\widetilde{\mathcal{M}}_3}$, is defined by
the formulation in~Sect.~\ref{sec:3} with the external gauge
field~\eqref{eq:(4.1)}, where~$\alpha=2\pi p/N$.
The integrand in Eq.~\eqref{eq:(4.35)} is manifestly invariant under the
physical $U(1)$ lattice gauge transformations and thus is well-defined. Also,
the defect~$\widetilde{M}_3$ is topological especially because the
sum~$\sum_{x\in\mathcal{M}_2^\nu}\epsilon_{\mu\nu\rho\sigma}f_{\rho\sigma}(x)$ is
invariant under deformations of~$\mathcal{M}_2^\nu\subset\mathcal{M}_3$ because
of the Bianchi identity.

Although our argument on the lattice correctly reproduces the
constraint~\eqref{eq:(4.33)} in continuum theory, something intriguing is
happening here: For~$\alpha=2\pi p/N$, the quantity in the square brackets
of~Eq.~\eqref{eq:(4.12)} would correspond to the Chern--Simons action
$-4ip/(16\pi N)\int_{\mathcal{M}_3}ada$. Under the gauge transformation,
$a\to a+d\phi$, this action naively changes
by~$-ip/(4\pi N)\int_{\mathcal{M}_3}d(\phi da)$. Then, setting
$\mathcal{M}_3=S^1\times\mathcal{M}_2$ and considering the winding~$k$
of~$\phi$ around~$S^1$, the change under the gauge transformation would
be~$-ipk/(2N)\int_{\mathcal{M}_2}da$. If this were correct, the sum over~$k$
would produce a constraint~$1/(2\pi)\int_{\mathcal{M}_2}da=2N\mathbb{Z}$ and this
does not coincide with~Eq.~\eqref{eq:(4.33)} by a factor~2. In~Appendix~A
of~Ref.~\cite{Karasik:2022kkq}, the necessity of the other factor~2 under the
gauge transformation is elucidated in detail. Here, it is interesting that our
lattice regularization provides this factor~2 automatically; the point is
that Eq.~\eqref{eq:(4.22)} is obtained from the gauge transformation on the
boundary variables which acts both of field strengths
in~$\epsilon_{\mu\nu\rho\sigma}f_{\mu\nu}(x)f_{\rho\sigma}(x+\Hat{\mu}+\Hat{\nu})$.

\section{Conclusion}
\label{sec:5}
In this paper, we made an attempt to realize the axial $U(1)$ noninvertible
symmetry of~Refs.~\cite{Cordova:2022ieu,Choi:2022jqy} in the framework of
lattice gauge theory. The structures of the axial $U(1)$ anomaly and the
associated Chern--Simons term with finite lattice spacings are controlled by
the lattice formulation of chiral gauge theory in~Ref.~\cite{Luscher:1998du}
based on the Ginsparg--Wilson relation with appropriate modifications for our
anomalous gauge theory. Imitating the prescription
of~Ref.~\cite{Karasik:2022kkq}, the symmetry operator/topological defect is
constructed by integrating the boundary variables along the defect
over smooth lattice gauge transformations. The projection operator for allowed
magnetic fluxes on the defect, with correct values, then emerges. The resulting
expression as the whole is manifestly invariant under the physical lattice
gauge transformations.

Our lattice formulation provides a firm basis for the axial $U(1)$
noninvertible symmetry in the $U(1)$ gauge theory and awaits possible
applications; the computation of the condensate and the fusion rule naturally
comes to mind.\footnote{See Appendix~\ref{sec:C}.} Also, a generalization to
the non-Abelian gauge theory is an intriguing problem. We hope to return to
these problems in the near future.

\section*{Acknowledgments}
We would like to thank
Motokazu Abe,
Yoshimasa Hidaka,
Naoto Kan,
Soichiro Shimamori,
Yuya Tanizaki,
Satoshi Yamaguchi,
and
Ryo Yokokura
for stimulating conversations that motivated the present work.
Our participation in the Yukawa Institute for Theoretical Physics at Kyoto
University (YITP) workshop ``Strings and Fields 2023'' (YITP-W-23-07) was also
quite valuable.
This work was partially supported by Japan Society for the Promotion of Science
(JSPS) Grant-in-Aid for Scientific Research Grant Numbers JP22KJ2096~(O.M.)
and~JP23K03418~(H.S.).

\appendix

\section{Counterterm in continuum theory}
\label{sec:A}
It is well-known that the definition of the gauge current requires close
attention when it possesses the anomaly~\cite{Bardeen:1984pm}. In this
Appendix, we illustrate this issue in continuum theory for our chiral gauge
theory with an anomalous matter content. This analysis also gives the important
information of what sort of counterterm should be included in our lattice
formulation in the main text.

We consider the left-handed Weyl fermion,
\begin{equation}
   P_-\psi=\psi,\qquad\Bar{\psi}P_+=\Bar{\psi}.
\label{eq:(A1)}
\end{equation}
The gauge group is $U(1)\times U(1)'$ and the charge assignment is the same
as that in the main text. Hence, we define the Dirac operator by
\begin{equation}
   \Slash{D}=\gamma_\mu D_\mu
   =\gamma_\mu(\partial_\mu+ia_\mu t+iA_\mu T),
\label{eq:(A2)}
\end{equation}
where representation matrices are
\begin{equation}
   t=\begin{pmatrix}
   +1&0\\0&-1\\
   \end{pmatrix},\qquad
   T=\begin{pmatrix}
   -1&0\\0&-1\\
   \end{pmatrix}.
\label{eq:(A3)}
\end{equation}
We define the effective action~${\mit\Gamma}$,
\begin{equation}
   e^{-{\mit\Gamma}}
   :=\int\mathrm{D}[\psi]\mathrm{D}[\Bar{\psi}]\,
   e^{-\int d^4x\,\Bar{\psi}\Slash{\scriptstyle{D}}\psi},
\label{eq:(A4)}
\end{equation}
that is a functional of $a_\mu$ and~$A_\mu$, in the following
way~\cite{Banerjee:1986bu}. First, we note
\begin{align}
   {\mit\Gamma}&=\int_0^1ds\,\frac{d}{ds}
   \left.{\mit\Gamma}\right|_{a_\mu\to sa_\mu,A_\mu\to sA_\mu}
\notag\\
   &=\int_0^1ds\,\frac{1}{s}\int d^4x\,\left.\left[
   a_\mu(x)\frac{\delta}{\delta a_\mu(x)}
   {\mit\Gamma}
   +A_\mu(x)\frac{\delta}{\delta A_\mu(x)}
   {\mit\Gamma}
   \right]\right|_{a_\mu\to sa_\mu,A_\mu\to sA_\mu}
\notag\\
   &=:i\int_0^1ds\,\frac{1}{s}\int d^4x\,\left.\left[
   a_\mu(x)\Tilde{\ell}_\mu(x)
   +A_\mu(x)\Tilde{L}_\mu(x)
   \right]\right|_{a_\mu\to sa_\mu,A_\mu\to sA_\mu},
\label{eq:(A5)}
\end{align}
where we have introduced gauge currents, $\Tilde{\ell}_\mu(x)$
and~$\Tilde{L}_\mu(x)$.

The representation~\eqref{eq:(A5)} of the effective action is, however, yet
formal because the gauge currents contain UV divergences. We hence
\emph{define} the gauge currents $\Tilde{\ell}_\mu(x)$ and~$\Tilde{L}_\mu(x)$
by the prescription~\cite{Fujikawa:1983bg},
\begin{align}
   \Tilde{l}_\mu(x)
   &=-\tr\left(
   P_+\gamma_\mu t\frac{1}{\Slash{D}}e^{\Slash{\scriptstyle{D}}^2/\Lambda^2}\right)
   \delta(x-y)|_{y\to x},
\notag\\
   \Tilde{L}_\mu(x)
   &=-\tr\left(
   P_+\gamma_\mu T\frac{1}{\Slash{D}}e^{\Slash{\scriptstyle{D}}^2/\Lambda^2}\right)
   \delta(x-y)|_{y\to x},
\label{eq:(A6)}
\end{align}
with the UV cutoff~$\Lambda$. Note that here the Dirac
operator~$\Slash{D}$~\eqref{eq:(A2)} does not contain the chiral projection
operator and the chirality projection is implemented by the insertion of~$P_+$
in the trace. This is a possible definition of the gauge current of a Weyl
fermion and this prescription is known as the covariant gauge
current~\cite{Fujikawa:1983bg}. Correspondingly, the way to define the
effective action through~Eqs.~\eqref{eq:(A5)} and~\eqref{eq:(A6)} is known as
the covariant regularization~\cite{Banerjee:1986bu}.
In~Ref.~\cite{Suzuki:1999qw}, it is shown that the lattice formulation
of~Ref.~\cite{Luscher:1998du}, at least in infinite volume, can be understood
in terms of the covariant regularization. In our present context, therefore, it
is natural to study the above prescription and resulting anomalies.

We can also define another sort of gauge current, the consistent gauge
current~\cite{Bardeen:1969md,Wess:1971yu}, as the functional derivative of the
effective action~${\mit\Gamma}$ with respect to the gauge potential.
From~Eq.~\eqref{eq:(A5)}, one has~\cite{Banerjee:1986bu} 
\begin{align}
   \ell_\mu(x)
   &=-i\frac{\delta}{\delta a_\mu(x)}{\mit\Gamma}
\notag\\
   &=\frac{\delta}{\delta a_\mu(x)}
   \int_0^1ds\,\frac{1}{s}\int d^4y\,\left.\left[
   a_\nu(y)\Tilde{\ell}_\nu(y)
   +A_\nu(y)\Tilde{L}_\nu(y)
   \right]\right|_{a_\mu\to sa_\mu,A_\mu\to sA_\mu}
\notag\\
   &=\int_0^1ds\,
   \left.\Tilde{\ell}_\mu(x)\right|_{a_\mu\to sa_\mu,A_\mu\to sA_\mu}
\notag\\
   &\qquad{}
   +\int_0^1ds\,\int d^4y\,\left.\left[
   a_\mu(y)\frac{\delta}{\delta a_\mu(x)}\Tilde{\ell}_\mu(y)
   +A_\mu(y)\frac{\delta}{\delta a_\mu(x)}\Tilde{L}_\mu(y)
   \right]\right|_{a_\mu\to sa_\mu,A_\mu\to sA_\mu}    
\notag\\
   &=\Tilde{\ell}_\mu(x)
   -\int_0^1ds\,s\frac{d}{ds}
   \left.\Tilde{\ell}_\mu(x)\right|_{a_\mu\to sa_\mu,A_\mu\to sA_\mu}
\notag\\
   &\qquad{}
   +\int_0^1ds\,\int d^4y\,\left.\left[
   a_\mu(y)\frac{\delta}{\delta a_\mu(x)}\Tilde{\ell}_\mu(y)
   +A_\mu(y)\frac{\delta}{\delta a_\mu(x)}\Tilde{L}_\mu(y)
   \right]\right|_{a_\mu\to sa_\mu,A_\mu\to sA_\mu}    
\notag\\
   &=\Tilde{\ell}_\mu(x)
   +\int_0^1ds\,\int d^4y\,\left.
   a_\nu(y)\left[
   \frac{\delta}{\delta a_\mu(x)}\Tilde{\ell}_\nu(y)
   -\frac{\delta}{\delta a_\nu(y)}\Tilde{\ell}_\mu(x)\right]
   \right|_{a_\mu\to sa_\mu,A_\mu\to sA_\mu}
\notag\\
   &\qquad{}
   +\int_0^1ds\,\int d^4y\,\left.
   A_\nu(y)\left[
   \frac{\delta}{\delta a_\mu(x)}\Tilde{L}_\nu(y)
   -\frac{\delta}{\delta A_\nu(y)}\Tilde{\ell}_\mu(x)\right]
   \right|_{a_\mu\to sa_\mu,A_\mu\to sA_\mu}.
\label{eq:(A7)}
\end{align}
In a similar way, we have
\begin{align}
   L_\mu(x)
   &=-i\frac{\delta}{\delta A_\mu(x)}{\mit\Gamma}
\notag\\
   &=\Tilde{L}_\mu(x)
   +\int_0^1ds\,\int d^4y\,\left.
   A_\nu(y)\left[
   \frac{\delta}{\delta A_\mu(x)}\Tilde{L}_\nu(y)
   -\frac{\delta}{\delta A_\nu(y)}\Tilde{L}_\mu(x)\right]
   \right|_{a_\mu\to sa_\mu,A_\mu\to sA_\mu}
\notag\\
   &\qquad{}
   +\int_0^1ds\,\int d^4y\,\left.
   a_\nu(y)\left[
   \frac{\delta}{\delta A_\mu(x)}\Tilde{\ell}_\nu(y)
   -\frac{\delta}{\delta a_\nu(y)}\Tilde{L}_\mu(x)\right]
   \right|_{a_\mu\to sa_\mu,A_\mu\to sA_\mu}.
\label{eq:(A8)}
\end{align}
Thus, the last two terms in~Eqs.~\eqref{eq:(A7)} and~\eqref{eq:(A8)} provide
the difference between the consistent current and the covariant current (this
difference is known as the Bardeen--Zumino current~\cite{Bardeen:1984pm}).

Starting from Eq.~\eqref{eq:(A6)}, one can then obtain explicit forms of the
last two terms in~Eqs.~\eqref{eq:(A7)} and~\eqref{eq:(A8)}. For details, see
Ref.~\cite{Banerjee:1986bu} and Sect.~6.6 of~Ref.~\cite{Fujikawa:2004cx}. After
some calculation, one has
\begin{align}
   &\int d^4y\,
   a_\nu(y)\left[
   \frac{\delta}{\delta a_\mu(x)}\Tilde{\ell}_\nu(y)
   -\frac{\delta}{\delta a_\nu(y)}\Tilde{\ell}_\mu(x)\right]
\notag\\
   &=i\int_0^1d\alpha\,\frac{1}{\Lambda^2}
   \tr\left[\gamma_5\gamma_\mu te^{(1-\alpha)\Slash{\scriptstyle{D}}^2/\Lambda^2}
   \Slash{a}te^{\alpha\Slash{\scriptstyle{D}}^2/\Lambda^2}\right]
   \delta(x-y)|_{y\to x}
\notag\\
   &\stackrel{\Lambda\to\infty}{=}
   -8\gamma\epsilon_{\mu\nu\rho\sigma}a_\nu(x)F_{\rho\sigma}(x),
\label{eq:(A9)}
\end{align}
where $\gamma=-1/(32\pi^2)$ and $\tr(t^3)=\tr(tT^2)=0$
and~$\tr(t^2T)=\tr(T^3)=-2$ have been noted. Similarly, we
find:\footnote{$\Lambda\to\infty$ is appropriately understood in what follows.}
\begin{align}
   \int d^4y\,
   A_\nu(y)\left[
   \frac{\delta}{\delta a_\mu(x)}\Tilde{L}_\nu(y)
   -\frac{\delta}{\delta A_\nu(y)}\Tilde{\ell}_\mu(x)\right]
   &=-8\gamma\epsilon_{\mu\nu\rho\sigma}A_\nu(x)f_{\rho\sigma}(x),
\notag\\
   \int d^4y\,
   A_\nu(y)\left[
   \frac{\delta}{\delta A_\mu(x)}\Tilde{L}_\nu(y)
   -\frac{\delta}{\delta A_\nu(y)}\Tilde{L}_\mu(x)\right]
   &=-8\gamma\epsilon_{\mu\nu\rho\sigma}A_\nu(x)F_{\rho\sigma}(x),
\notag\\
   \int d^4y\,
   a_\nu(y)\left[
   \frac{\delta}{\delta A_\mu(x)}\Tilde{\ell}_\nu(y)
   -\frac{\delta}{\delta a_\nu(y)}\Tilde{L}_\mu(x)\right]
   &=-8\gamma\epsilon_{\mu\nu\rho\sigma}a_\nu(x)f_{\rho\sigma}(x).
\label{eq:(A10)}
\end{align}
Plugging these into~Eqs.~\eqref{eq:(A7)} and~\eqref{eq:(A8)} and noting
$\int_0^1ds\,s^2=1/3$, we have the relation between the consistent and
covariant gauge currents,
\begin{align}
   \ell_\mu(x)&=\Tilde{\ell}_\mu(x)
   -\frac{8}{3}\gamma\epsilon_{\mu\nu\rho\sigma}
   \left[a_\nu(x)F_{\rho\sigma}(x)+A_\nu(x)f_{\rho\sigma}(x)\right],
\notag\\
   L_\mu(x)&=\Tilde{L}_\mu(x)
   -\frac{8}{3}\gamma\epsilon_{\mu\nu\rho\sigma}
   \left[a_\nu(x)f_{\rho\sigma}(x)+A_\nu(x)F_{\rho\sigma}(x)\right].
\label{eq:(A11)}
\end{align}
On the other hand, the covariant gauge currents~\eqref{eq:(A6)} give rise to
the so-called covariant gauge anomaly~\cite{Fujikawa:1983bg}:
\begin{align}
   \partial_\mu\Tilde{\ell}_\mu(x)
   &=2\gamma\epsilon_{\mu\nu\rho\sigma}
   \left[f_{\mu\nu}(x)F_{\rho\sigma}(x)+F_{\mu\nu}(x)f_{\rho\sigma}(x)\right],
\notag\\
   \partial_\mu\Tilde{L}_\mu(x)
   &=2\gamma\epsilon_{\mu\nu\rho\sigma}
   \left[f_{\mu\nu}(x)f_{\rho\sigma}(x)+F_{\mu\nu}(x)F_{\rho\sigma}(x)\right].   
\label{eq:(A12)}
\end{align}
The anomaly of the consistent gauge currents~\eqref{eq:(A11)}, the so-called
consistent gauge anomaly~\cite{Bardeen:1969md,Wess:1971yu}, is therefore given
by
\begin{align}
   \partial_\mu\ell_\mu(x)
   &=\frac{2}{3}\gamma\epsilon_{\mu\nu\rho\sigma}
   \left[f_{\mu\nu}(x)F_{\rho\sigma}(x)+F_{\mu\nu}(x)f_{\rho\sigma}(x)\right],
\notag\\
   \partial_\mu L_\mu(x)
   &=\frac{2}{3}\gamma\epsilon_{\mu\nu\rho\sigma}
   \left[f_{\mu\nu}(x)f_{\rho\sigma}(x)+F_{\mu\nu}(x)F_{\rho\sigma}(x)\right].   
\label{eq:(A13)}
\end{align}

From Eq.~\eqref{eq:(A13)}, we observe that in the present prescription (i.e.,
in the covariant regularization), (i)~$\ell_\mu(x)$, which corresponds to the
vector current in the target theory, does not conserve; and (ii)~the
coefficient of~$\epsilon_{\mu\nu\rho\sigma}f_{\mu\nu}f_{\rho\sigma}$
in~$\partial_\mu L_\mu(x)$, which corresponds to the axial vector anomaly in the
target theory, is $1/3$ of the naively expected one in~Eq.~\eqref{eq:(A12)}; if
one computes the divergence of the axial vector current while respecting the
vectorial gauge invariance, one has the coefficient
of~$\epsilon_{\mu\nu\rho\sigma}f_{\mu\nu}f_{\rho\sigma}$
in~$\partial_\mu\Tilde{L}_\mu(x)$ of~Eq.~\eqref{eq:(A12)}. These two facts are
actually related to each other. One can make $\ell_\mu(x)$ conserving by adding
an appropriate counterterm to the effective action. The required counterterm is
\begin{equation}
   -i\Delta{\mit\Gamma}=\frac{8}{3}\gamma
   \int d^4x\,\epsilon_{\mu\nu\rho\sigma}A_\mu(x)a_\nu(x)f_{\rho\sigma}(x).
\label{eq:(A14)}
\end{equation}
This provides the additional contribution to the consistent gauge anomalies
and
\begin{align}
   \partial_\mu\left[
   \ell_\mu(x)-i\frac{\delta}{\delta a_\mu(x)}\Delta{\mit\Gamma}\right]&=0,
\notag\\
   \partial_\mu\left[
   L_\mu(x)-i\frac{\delta}{\delta A_\mu(x)}\Delta{\mit\Gamma}\right]
   &=2\gamma\epsilon_{\mu\nu\rho\sigma}f_{\mu\nu}(x)f_{\rho\sigma}(x)
   +\frac{2}{3}\gamma\epsilon_{\mu\nu\rho\sigma}F_{\mu\nu}(x)F_{\rho\sigma}(x).
\label{eq:(A15)}
\end{align}
This is the desired form of the consistent gauge anomalies in our context. That
is, the vector current conserves and the
$\gamma\epsilon_{\mu\nu\rho\sigma}f_{\mu\nu}f_{\rho\sigma}$ term in the axial
anomaly has the naively expected coefficient. We want to realize this structure
also in the lattice formulation in the main text.

Equation~\eqref{eq:(A14)} indicates what sort of counterterm should be added in
the effective action in our lattice formulation. In the measure
term~\eqref{eq:(3.48)}, the term being proportional to~$\gamma$ is the
variation of a lattice counterpart of the counterterm~\eqref{eq:(A14)}. This
term renders the coefficient of the
$\epsilon_{\mu\nu\rho\sigma}f_{\mu\nu}f_{\rho\sigma}$ term in the axial anomaly
the naively expected one in~Eq.~\eqref{eq:(3.53)}.

\section{Gauge integration measure}
\label{sec:B}
In this Appendix, we give a precise definition of the integration measure for
the smooth gauge degrees of freedom assumed in the main text, e.g.,
in~Eqs.~\eqref{eq:(4.12)} and~\eqref{eq:(4.27)}.

We define the measure by
\begin{equation}
   \int\mathrm{D}[\lambda]
   =\prod_{x\in\mathcal{M}_3}
   \left[\int_{-\pi}^\pi\frac{\mathrm{d}\phi(x)}{2\pi}\right]
   W[\phi],
\label{eq:(B1)}
\end{equation}
where
\begin{equation}
   W[\phi]^{-1}
   :=w[\phi]^{-1}\prod_{x\in\mathcal{M}_3}
   \left[\int_{-\pi}^\pi\frac{\mathrm{d}\phi'(x)}{2\pi}\right]
   \prod_{\nu\neq\mu}\prod_{y\in\mathcal{M}_3-\mathcal{M}_2^\nu}
   \delta(\partial\phi(y,\nu)-\partial\phi'(y,\nu)),
\label{eq:(B2)}
\end{equation}
and
\begin{align}
   w[\phi]^{-1}
   &=\prod_{\nu\neq\mu}
   \prod_{x\in\mathcal{M}_3}
   \left[
   \int_{-\epsilon'}^{\epsilon'}d\theta_\nu(x)\right]
   \prod_{y\in\mathcal{M}_2^\nu}
   \delta\left(
   \sum_{t=0}^{\ell-1}\theta_\nu(y+t\Hat{\nu})
   -\sum_{t=0}^{\ell-1}\partial\phi(y+t\Hat{\nu},\nu)\right)
\notag\\
   &=\prod_{\nu\neq\mu}\ell^2\int_{-\infty}^\infty\frac{d\alpha}{2\pi}\,
   e^{-i\sum_{t=0}^{\ell-1}\partial\phi(x+t\Hat{\nu},\nu)\alpha}
   \left[\frac{2}{\alpha}\sin(\epsilon'\alpha)\right]^\ell.
\label{eq:(B3)}
\end{align}
In these expressions, we have introduced the symbol (see Eq.~\eqref{eq:(4.16)})
\begin{equation}
   \partial\phi(x,\nu)
   :=\frac{1}{i}\ln\left[e^{-i\phi(x)}e^{i\phi(x+\Hat{\nu})}\right]
   =\partial_\nu\phi(x)+2\pi l_\nu(x)
\label{eq:(B4)}
\end{equation}
and $\mathcal{M}_2^\nu$ denotes the submanifold such that
$\mathcal{M}_3=S^1\times\mathcal{M}_2^\nu$, where $S^1$ being the direction
of~$\nu$.

Now, noting the smoothness in~Eq.~\eqref{eq:(4.17)},
$|\partial\phi(x,\nu)|<\epsilon'$, and that
$\frac{1}{2\pi}\sum_{t=0}^{\ell-1}\partial\phi(x+t\Hat{\nu})$ is an
integer as~Eq.~\eqref{eq:(4.24)}, we have
\begin{equation}
   \prod_{\nu\neq\mu}
   \left[\sum_{k_\nu=-[\epsilon'\ell/(2\pi)]}^{[\epsilon'\ell/(2\pi)]}
   \delta_{k_\nu,
   \frac{1}{2\pi}\sum_{t=0}^{\ell-1}\partial\phi(x+t\Hat{\nu},\nu)}\right]
   \prod_{\rho\neq\mu}\prod_{y\in\mathcal{M}_3}
   \left[
   \int_{-\epsilon'}^{\epsilon'}d\theta_\rho(y)\,
   \delta(\theta_\rho(y)-\partial\phi(y,\rho))\right]=1,
\label{eq:(B5)}
\end{equation}
where $x\in\mathcal{M}_3$ is arbitrary, and we insert this
into~Eq.~\eqref{eq:(B1)}. After some consideration, we then have
\begin{align}
   &\int\mathrm{D}[\lambda]\,F[\phi]
\notag\\
   &=
   \prod_{\nu\neq\mu}
   \left[\sum_{k_\nu=-[\epsilon'\ell/(2\pi)]}^{[\epsilon'\ell/(2\pi)]}\right]
   \prod_{\rho\neq\mu}\prod_{x\in\mathcal{M}_3}
   \left[
   \int_{-\epsilon'}^{\epsilon'}d\theta_\rho(x)\right]
   \prod_{y\in\mathcal{M}_2^\rho}
   \delta\left(
   \sum_{t=0}^{\ell-1}\theta_\rho(y+t\Hat{\rho})-2\pi k_\rho
   \right)w[\phi]
   F[\phi],
\label{eq:(B6)}
\end{align}
where, in the integrand~$F[\phi]$, we have the replacement,\footnote{Here,
we assume that $F[\phi]$ is invariant under the global
shift~$\phi(x)\to\phi(x)+\theta$. Otherwise, we have another integration
over~$\phi(o)$, $\int_{-\pi}^\pi\frac{d\phi(o)}{2\pi}$, where
$o\in\mathcal{M}_3$ is a certain fixed point.}
\begin{equation}
   \partial\phi(x,\nu)\to\theta_\nu(x).
\label{eq:(B7)}
\end{equation}
When the integrand~$F[\phi]$ is a function of the winding numbers~$k_\nu$ only,
we can carry out the integration over $\theta_\rho(x)$ in~Eq.~\eqref{eq:(B6)}
and it cancels the factor~$w[\phi]$ in~Eq.~\eqref{eq:(B3)}. Therefore, in such
a case, the measure reduces to a simple sum over winding numbers,
\begin{equation}
   \int\mathrm{D}[\lambda]\,F[\phi]
   =\prod_{\nu\neq\mu}
   \left[\sum_{k_\nu=-[\epsilon'\ell/(2\pi)]}^{[\epsilon'\ell/(2\pi)]}\right]
   F[\phi],
\label{eq:(B8)}
\end{equation}
as we assume in~Eq.~\eqref{eq:(4.27)}.

\section{Symmetry operator in terms of a $\mathbb{Z}_N$ TQFT}
\label{sec:C}
In the construction of the symmetry operator in the main text, some points
remain unsatisfactory. One is that the construction of the symmetry operator is
limited to cases of homologically trivial 3-dimensional (3D) closed surfaces,
because the construction refers to an auxiliary 4-dimensional (4D) volume on
the lattice. This is unsatisfactory from the perspective of the construction of
the symmetry operator for generic cases and possible applications. The another
point is that the invariance of the symmetry operator under the gauge
transformation along the 3D surface is limited to sufficiently smooth lattice
gauge transformations. These two points are mutually related and have the same
root. In this Appendix, we provide an intrinsically 3D construction of the
symmetry operator by employing a 3D $\mathbb{Z}_N$ TQFT, the level-$N$ BF
theory, whose properties have been well-understood~\cite{Banks:2010zn,Kapustin:2014gua,Kaidi:2021gbs,Choi:2021kmx,Schafer-Nameki:2023jdn,Shao:2023gho}. In
particular, we give the construction of the symmetry operator for generic 3D
closed surfaces. The 3D expression is moreover manifestly invariant under
\emph{arbitrary\/} 3D lattice gauge transformations. As an application of our
construction of the symmetry operator, we give the evaluation of fusion rules
of symmetry operators within our lattice regularized framework.

In what follows, we set the rotation angle as
\begin{equation}
   \alpha=\frac{2\pi p}{N}
\label{eq:(C1)}
\end{equation}
by integers $p$ and~$N$. We first assume that $p$ and~$N$ are coprime and later
relax this restriction. Our formulation on the basis of the level-$N$ BF theory
is possible only when $p$ is an even integer~$p=2\mathbb{Z}$.

First, using Eq.~\eqref{eq:(4.11)}, Eq.~\eqref{eq:(4.10)} is written as
\begin{align}
   &\langle\mathcal{O}\rangle_{\mathrm{F}}^{\widetilde{\mathcal{M}}_3}
   \exp
   \left\{
   -\frac{ip}{8\pi N}\sum_{x\in\mathcal{M}_3}
   \epsilon_{\mu\nu\rho}
   \left[a_\mu(x)f_{\nu\rho}(x+\Hat{\mu})
   +2\pi z_{\mu\nu}(x)a_\rho(x+\Hat{\mu}+\Hat{\nu})\right]\right\}
\notag\\
   &\qquad{}
   \times\exp
   \left[
   -\frac{ip\pi}{4N}\sum_{x\in\mathcal{V}_4}
   \epsilon_{\mu\nu\rho\sigma}
   z_{\mu\nu}(x)z_{\rho\sigma}(x+\Hat{\mu}+\Hat{\nu})
   \right]
   =\langle\mathcal{O}^\alpha\rangle_{\mathrm{F}}.
\label{eq:(C2)}
\end{align}
In terms of the cochain and the cup product on the hypercubic
lattice~\cite{Chen:2021ppt,Jacobson:2023cmr},\footnote{These notions have also
been known in the context of the non-commutative differential
calculus~\cite{Fujiwara:1999fi}.} this can also be written as
\begin{equation}
   \langle\mathcal{O}\rangle_{\mathrm{F}}^{\widetilde{\mathcal{M}}_3}
   \exp
   \left[
   -\frac{ip}{4\pi N}\sum_{\mathrm{cube}\in\mathcal{M}_3}
   \left(a\cup f+2\pi z\cup a\right)\right]
   \exp
   \left(
   -\frac{ip\pi}{N}\sum_{\mathrm{hypercube}\in\mathcal{V}_4}z\cup z
   \right)
   =\langle\mathcal{O}^\alpha\rangle_{\mathrm{F}},
\label{eq:(C3)}
\end{equation}
where $a$ is a 1-cochain in~$\Gamma$, $a\in C^1(\Gamma;\mathbb{R})$, $z$ is a
$\mathbb{Z}$ 2-cocycle in~$\Gamma$ (because of Eq.~\eqref{eq:(2.15)},
$\delta z=0$, where $\delta$ is the coboundary operator),
$z\in Z^2(\Gamma;\mathbb{Z})$, and~$f$ is a 2-cocycle,
$f=\delta a+2\pi z\in Z^2(\Gamma;\mathbb{R})$.

In the anomalous Ward--Takahashi identity~\eqref{eq:(C3)}, the first
exponential on the left-hand side has the form of a 3D sum and this part can be
generalized to arbitrary, i.e.\ not necessarily homologically trivial, closed
3D surfaces. The problem is the second exponential because it refers to a 4D
volume~$\mathcal{V}_4$.

\subsection{Symmetry operator in terms of the BF theory on the lattice}
\label{sec:C.1}
\begin{figure}[htbp]
\centering
\begin{tikzpicture}[scale=0.7]
  \draw[blue,ultra thick,dotted] (0,0)--(1,0);
  \draw[blue,ultra thick] (1,0)--(2,0) node[below] {$\Tilde{x}+\Hat{\mu}$};
  \draw[very thick] (-1,-1)--(1,-3)--(1,1)--(-1,3)--cycle;
  \draw[very thick] (-1,-1)--(-5,-1);
  \node[below] at (1,-3) {$x+\Hat{\mu}+\Hat{\nu}$};
  \node[above] at (-1,3) {$x+\Hat{\mu}+\Hat{\rho}$};
  \node[below] at (-5,-1) {$x$};
  \node[below left] at (-1,-1) {$x+\Hat{\mu}$};
  \draw[blue,ultra thick] (0,0)--(-2,0) node[below] {$\Tilde{x}$};
  \draw[->,thick] (-9,-2)--(-8,-2) node[right] {$\mu$};
  \draw[->,thick] (-9,-2)--(-9,-1) node[above] {$\rho$};
  \draw[->,thick] (-9,-2)--(-9+0.707,-2-0.707) node[below] {$\nu$};
  \fill[blue] (-2,0) circle(3pt);
  \fill[blue] (2,0) circle(3pt);
  \fill (-1,-1) circle(3pt);
  \fill (1,-3) circle(3pt);
  \fill (1,1) circle(3pt);
  \fill (-1,3) circle(3pt);
  \fill (-5,-1) circle(3pt);
\end{tikzpicture}
\caption{The structure of the product in the first term of the lattice
action~\eqref{eq:(C4)}. The 1-cochain $b$ is put on the dual link (the blue
line) and the 1-cochain~$c$ is put on the original lattice (the black lines);
2-cochain $\delta c-z$ is therefore put on the original plaquette (the
square).}
\label{fig:C1}
\end{figure}
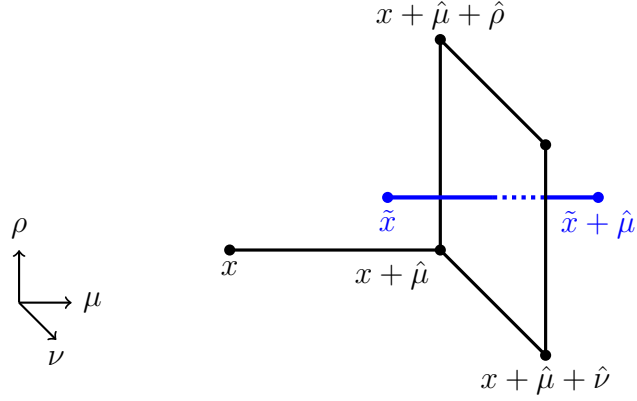

To express that second exponential in~Eq.~\eqref{eq:(C3)} solely in terms of 3D
notions, we introduce the level-$N$ BF theory defined on a 3-dimensional closed
surface (i.e.\ 3-cycle) $\mathcal{M}_3$ in~$\Gamma$. We define the action on a
cubic lattice by
\begin{align}
   S_{\mathrm{BF}}
   &=\frac{-ip\pi}{N}\sum_{\mathrm{cube}\in\mathcal{M}_3}
   \left[
   b(\delta c-z)-z\cup c
   \right]
\notag\\
   &=\frac{-ip\pi}{N}
   \sum_{x\in\mathcal{M}_3}
   \epsilon_{\mu\nu\rho}
   \left\{
   b_\mu(\Tilde{x})\left[\partial_\nu c_\rho(x+\Hat{\mu})
   -\frac{1}{2}z_{\nu\rho}(x+\Hat{\mu})\right]
   -\frac{1}{2}z_{\mu\nu}(x)c_\rho(x+\Hat{\mu}+\Hat{\nu})
   \right\},
\label{eq:(C4)}
\end{align}
where $b\in C^1(\widetilde{\mathcal{M}}_3;\mathbb{Z}_N)$
and~$c\in C^1(\mathcal{M}_3;\mathbb{Z}_N)$; $z$ is a background 2-cocycle
$z\in Z^2(\mathcal{M}_3;\mathbb{Z})$, $\delta z=0$. In the lattice
action~\eqref{eq:(C4)}, we put the $\mathbb{Z}_N$ 1-cochain~$b$ on the dual
lattice, whose sites are given by
\begin{equation}
   \Tilde{x}=x+\frac{1}{2}\Hat{\mu}+\frac{1}{2}\Hat{\nu}+\frac{1}{2}\Hat{\rho},
\label{eq:(C5)}
\end{equation}
assuming that $\mathcal{M}_3$ extends in $\mu$, $\nu$, $\rho$ directions. The
first term of the action~\eqref{eq:(C4)} has the structure depicted
in~Fig.~\ref{fig:C1}; note that this is \emph{not\/} the cup product.

The partition function of the BF theory is then defined by
\begin{equation}
   \mathcal{Z}_{\mathcal{M}_3}[z]
   =\frac{1}{|C^0(\mathcal{M}_3;\mathbb{Z}_N)|}
   \int\mathrm{D}[b]\mathrm{D}[c]\,
   e^{-S_{\mathrm{BF}}},
\label{eq:(C6)}
\end{equation}
where
\begin{equation}
   \mathrm{D}[b]:=
   \prod_{\mu,x\in\mathcal{M}_3}
   \left[\frac{1}{N}\sum_{b_\mu(\Tilde{x})=0}^{N-1}\right],\qquad
   \mathrm{D}[c]:=
   \prod_{\mu,x\in\mathcal{M}_3}\left[\sum_{c_\mu(x)=0}^{N-1}\right].
\label{eq:(C7)}
\end{equation}
Note that for~$p$ even $e^{-S_{\mathrm{BF}}}$ is invariant under
$b_\mu(\Tilde{x})\to b_\mu(\Tilde{x})+N\mathbb{Z}$
and~$c_\mu(x)\to c_\mu(x)+N\mathbb{Z}$ and this is consistent with the summation
range in~Eq.~\eqref{eq:(C7)}. Also, since $\delta z=0$, $e^{-S_{\mathrm{BF}}}$ is
invariant under the 0-form $\mathbb{Z}_N$ gauge transformation,
$c\to c+\delta\mu$, where $\mu\in C^0(\mathcal{M}_3;\mathbb{Z}_N)$. Therefore,
it is natural to define the partition function~\eqref{eq:(C6)} by dividing it
by the gauge volume~$|C^0(\mathcal{M}_3;\mathbb{Z}_N)|=N^s$, where $s$ denotes
the number of sites in~$\mathcal{M}_3$.

Now, taking the sum over~$b$ in Eq.~\eqref{eq:(C6)}, for $p$ even, we have
\begin{equation}
   \mathcal{Z}_{\mathcal{M}_3}[z]
   =\frac{1}{|C^0(\mathcal{M}_3;\mathbb{Z}_N)|}
   \int\mathrm{D}[c]\,
   \delta_N\left[\delta c-z\right]
   \exp\left(
   -\frac{ip\pi}{N}
   \sum_{\mathrm{cube}\in\mathcal{M}_3}z\cup c\right),
\label{eq:(C8)}
\end{equation}
where we have introduced the delta functional:
\begin{equation}
   \delta_N\left[\delta c-z\right]
   :=\prod_{\mu,x\in\mathcal{M}_3,\nu\neq\mu,\rho\neq\mu,\nu<\rho}
   \delta_{
   \partial_\nu c_\rho(x+\Hat{\mu})-\partial_\rho c_\nu(x+\Hat{\mu})
   -z_{\nu\rho}(x+\Hat{\mu}),N\mathbb{Z}}.
\label{eq:(C9)}
\end{equation}
Equation~\eqref{eq:(C8)} immediately shows that, if there exists a 2D closed
surface~$\mathcal{M}_2\subset\mathcal{M}_3$ such that the magnetic flux is
nonzero modulo~$N$, i.e.
\begin{equation}
   \sum_{\mathrm{face}\in\mathcal{M}_2}z
   =\frac{1}{2}\sum_{x\in\mathcal{M}_2}\epsilon_{\mu\nu}z_{\mu\nu}(x)
   \neq0\bmod N,
\label{eq:(C10)}
\end{equation}
then there is no configuration of~$c$ such that $\delta c=z\bmod N$. In this
case, therefore, the partition function identically vanishes,
$\mathcal{Z}_{\mathcal{M}_3}[z]=0$. Later, we substitute $z\to z|_{\mathcal{M}_3}$,
where $z|_{\mathcal{M}_3}$ denotes the 2-cocycle $z\in Z^2(\Gamma,\mathbb{Z})$
in~Eq.~\eqref{eq:(C3)} restricted on~$\mathcal{M}_3$. If this~$z|_{\mathcal{M}_3}$
satisfies~Eq.~\eqref{eq:(C10)} (this is equivalent
to~$[1/(2\pi)]\sum_{\mathrm{face}\in\mathcal{M}_2}f\neq0\bmod N$), we cannot
construct the symmetry operator. The existence of the symmetry operator,
therefore, requires that the magnetic fluxes in~Eq.~\eqref{eq:(2.11)} are
quantized in unit of~$N$.

Let us first set~$z=0$ in~Eq.~\eqref{eq:(C8)}. Then, letting numbers of the
sites, links, faces (plaquettes), and cubes of~$\mathcal{M}_3$, be $s$, $l$,
$f$, and~$c$, respectively, the partition function is evaluated as
\begin{align}
   \mathcal{Z}_{\mathcal{M}_3}[0]
   &=\frac{1}{|C^0(\mathcal{M}_3;\mathbb{Z}_N)|}
   \int\mathrm{D}[c]\,
   \delta_N\left[\delta c\right]
\notag\\
   &=\frac{1}{N^s}N^l\frac{1}{N^{f-c+1}}N^{b_2}
\notag\\
   &=N^{b_2-1},
\label{eq:(C11)}
\end{align}
where $b_2=|H^2(\mathcal{M}_3;\mathbb{Z}_N)|$ is the second Betti number
of~$\mathcal{M}_3$. In the second line, various powers of~$N$ have the
following origins: First, $|C^0(\mathcal{M}_3;\mathbb{Z}_N)|=N^s$. The number
of~$c$ is~$l$ and we have $N^l$ from the sum over~$c$. This counting is of
course too many because there is a constraint~$\delta c=0$ for each face. The
number of independent constraints is~$f-c+1$ because the constraints on five
faces of a cube imply the constraint in the last face; on a closed surface,
however, a one cube does not contribute to the reduction of the constraints.
These explain the factor~$1/N^{f-c+1}$. The last factor $N^{b_2}$ comes from the
number of independent nontrivial cycles on $\mathcal{M}_3$, i.e.,
$b_2=|H^2(\mathcal{M}_3;\mathbb{Z}_N)|$. This is the number of the
cohomologically nontrivial solutions of~$\delta c=0\bmod N$. In the last
equality, we have noted that the Euler number vanishes for 3-cycles,
$s-l+f-c=\chi(\mathcal{M}_3)=0$. Therefore, with the definition of the
integration measure in~Eq.~\eqref{eq:(C7)}, the partition
function~$\mathcal{Z}_{\mathcal{M}_3}[0]$ is topological, depending only on the
topology of~$\mathcal{M}_3$. For instance, we have
\begin{equation}
   \mathcal{Z}_{S^3}[0]=\frac{1}{N},\qquad
   \mathcal{Z}_{S^2\times S^1}[0]=1,\qquad
   \mathcal{Z}_{T^3}[0]=N^2.
\label{eq:(C12)}
\end{equation}

Next, let us consider the case of~$z\neq0$. As noted in~Eq.~\eqref{eq:(C10)},
if there exists a 2-cycle $\mathcal{M}_2\subset\mathcal{M}_3$, such that
$\sum_{\mathrm{face}\in\mathcal{M}_2}z\neq0\bmod N$, the partition function
identically vanishes, $\mathcal{Z}_{\mathcal{M}_3}[z]=0$. In other words, this
occurs if $z\bmod N\neq0$ in~$H^2(\mathcal{M}_3;\mathbb{Z}_N)$. Therefore, let
us assume that $z\bmod N=0$ in~$H^2(\mathcal{M}_3;\mathbb{Z}_N)$. Then, there
exists $\nu\in C^1(\mathcal{M}_3;\mathbb{Z}_N)$ such that
\begin{equation}
   z=\delta\nu\bmod N.
\label{eq:(C13)}
\end{equation}
Substituting this into~Eq.~\eqref{eq:(C8)}, we have
\begin{align}
   \mathcal{Z}_{\mathcal{M}_3}[z]
   &=\frac{1}{|C^0(\mathcal{M}_3;\mathbb{Z}_N)|}
   \int\mathrm{D}[c]\,
   \delta_N\left[\delta c-\delta\nu\right]
   \exp\left(
   -\frac{ip\pi}{N}
   \sum_{\mathrm{cube}\in\mathcal{M}_3}\delta\nu\cup c\right)
\notag\\
   &=\exp\left(
   -\frac{ip\pi}{N}
   \sum_{\mathrm{cube}\in\mathcal{M}_3}\delta\nu\cup\nu\right)
   \mathcal{Z}_{\mathcal{M}_3}[0],
\label{eq:(C14)}
\end{align}
when $z\bmod N=0$ in~$H^2(\mathcal{M}_3;\mathbb{Z}_N)$. Here, we have first
shifted $c$ as $c\to c+\nu\bmod N$ and then noted by the Leibniz rule,
$\delta\nu\cup c=\delta(\nu\cup c)+\nu\cup\delta c$. Finally, we have used
$\delta c=0\bmod N$ which holds under the delta functional after the shift.

We now set $z\to z|_{\mathcal{M}_3}$, where $z|_{\mathcal{M}_3}$ is the 2-cocycle
$z\in Z^2(\Gamma,\mathbb{Z})$ in~Eq.~\eqref{eq:(C3)} restricted
to~$\mathcal{M}_3$. We then consider a ``small'' deformation of the
3-cycle~$\mathcal{M}_3\to\mathcal{M}_3'$ within~$\Gamma$. Here, by small, we
mean that there exists a 4-volume (ball)~$\mathcal{V}_4$ such that
$\partial(\mathcal{V}_4)=\mathcal{M}_3'\cup(-\mathcal{M}_3)$. Since
$\delta z=0$ on~$\Gamma$, under such a small deformation,
\begin{equation}
   \text{$z|_{\mathcal{M}_3}\bmod N\neq0$ in~$H^2(\mathcal{M}_3;\mathbb{Z}_N)$}
   \Leftrightarrow
   \text{$z|_{\mathcal{M}_3'}\bmod N\neq0$ in~$H^2(\mathcal{M}_3';\mathbb{Z}_N)$},
\label{eq:(C15)}
\end{equation}
and, for such cases, as noted in~Eq.~\eqref{eq:(C10)},
\begin{equation}
   \mathcal{Z}_{\mathcal{M}_3}[z|_{\mathcal{M}_3}]
   =\mathcal{Z}_{\mathcal{M}_3'}[z|_{\mathcal{M}_3'}]=0.
\label{eq:(C16)}
\end{equation}
Therefore, let us assume that
$z|_{\mathcal{M}_3}\bmod N=0$ in~$H^2(\mathcal{M}_3;\mathbb{Z}_N)$
(and thus $z|_{\mathcal{M}_3'}\bmod N=0$ in~$H^2(\mathcal{M}_3';\mathbb{Z}_N)$).
This implies that there exist $\nu\in H^1(\mathcal{M}_3;\mathbb{Z}_N)$
such that $z|_{\mathcal{M}_3}=\delta\nu$
and $\nu'\in H^1(\mathcal{M}_3';\mathbb{Z}_N)$
such that $z|_{\mathcal{M}_3'}=\delta\nu'$. Therefore,
from~Eqs.~\eqref{eq:(C14)} and~\eqref{eq:(C11)},
\begin{align}
   &\mathcal{Z}_{\mathcal{M}_3'}[z|_{\mathcal{M}_3'}]
\notag\\
   &=\exp\left(
   -\frac{ip\pi}{N}
   \sum_{\mathrm{cube}\in\mathcal{M}_3'}\delta\nu'\cup\nu'\right)
   \mathcal{Z}_{\mathcal{M}_3'}[0]
\notag\\
   &=\exp\left[
   -\frac{ip\pi}{N}
   \left(
   \sum_{\mathrm{cube}\in\mathcal{M}_3'}\delta\nu'\cup\nu'
   -\sum_{\mathrm{cube}\in\mathcal{M}_3}\delta\nu\cup\nu
   \right)
   \right]
   \mathcal{Z}_{\mathcal{M}_3}[z|_{\mathcal{M}_3}].
\label{eq:(C17)}
\end{align}
Now, since $\mathcal{V}_4$ is a ball, $H^2(\mathcal{V}_4,\mathbb{Z})=0$ and,
for $\delta z=0$, there exists $\Tilde{\nu}$ such that
\begin{equation}
   z|_{\mathcal{V}_4}=\delta \Tilde{\nu},\qquad
   \Tilde{\nu}\in C^1(\mathcal{V}_4,\mathbb{Z}).
\label{eq:(C18)}
\end{equation}
This implies that
\begin{align}
   \sum_{\mathrm{hypercube}\in\mathcal{V}_4}
   z\cup z
   &=\sum_{\mathrm{hypercube}\in\mathcal{V}_4}
   \delta(\delta\Tilde{\nu}\cup\Tilde{\nu})
\notag\\
   &=\sum_{\mathrm{cube}\in\mathcal{M}_3'}
   \delta\nu'\cup\nu'
   -\sum_{\mathrm{cube}\in\mathcal{M}_3}
   \delta\nu\cup\nu\bmod N
\label{eq:(C19)}
\end{align}
and Eq.~\eqref{eq:(C17)} shows
\begin{equation}
   \mathcal{Z}_{\mathcal{M}_3'}[z|_{\mathcal{M}_3'}]
   =\exp\left(
   -\frac{ip\pi}{N}
   \sum_{\mathrm{hypercube}\in\mathcal{V}_4}
   z\cup z
   \right)
   \mathcal{Z}_{\mathcal{M}_3}[z|_{\mathcal{M}_3}].
\label{eq:(C20)}
\end{equation}
It is clear that this relation holds as far as 3-cycles $\mathcal{M}_3'$
and~$\mathcal{M}_3$ are related by repeated applications of the small
deformations; in such a case,
$\partial(\mathcal{V}_4)=\mathcal{M}_3'\cup(-\mathcal{M}_3)$.
Comparing this with~Eq.~\eqref{eq:(C3)}, we arrive at the conclusion that the
symmetry operator on an \emph{arbitrary\/} 3-cycle~$\widetilde{M}_3$ is given
by using the BF theory as
\begin{equation}
   \left\langle
   \Tilde{U}_{2\pi p/N}(\widetilde{M}_3)\mathcal{O}
   \right\rangle_{\mathrm{F}}
   =\left\langle
   \mathcal{O}
   \right\rangle_{\mathrm{F}}^{\widetilde{M}_3}
   \exp
   \left[
   -\frac{ip}{4\pi N}\sum_{\mathrm{cube}\in\mathcal{M}_3}
   \left(a\cup f+2\pi z\cup a\right)\right]
   \mathcal{Z}_{\mathcal{M}_3}[z|_{\mathcal{M}_3}].
\label{eq:(C21)}
\end{equation}
If the deformation of the defect~$\widetilde{M}_3\to\widetilde{M}_3'$ can be
realized as above, from~Eq.~\eqref{eq:(C20)},
\begin{align}
   &\left\langle
   \Tilde{U}_{2\pi p/N}(\widetilde{M}_3')\mathcal{O}
   \right\rangle_{\mathrm{F}}
\notag\\
   &=\left\langle
   \mathcal{O}
   \right\rangle_{\mathrm{F}}^{\widetilde{M}_3'}
   \exp
   \left[
   -\frac{ip}{4\pi N}\sum_{\mathrm{hypercube}\in\mathcal{V}_4}
   \delta\left(a\cup f+2\pi z\cup a\right)\right]
   \exp\left(
   -\frac{ip\pi}{N}
   \sum_{\mathrm{hypercube}\in\mathcal{V}_4}
   z\cup z
   \right)
\notag\\
   &\qquad\qquad{}
   \times\exp
   \left[
   -\frac{ip}{4\pi N}\sum_{\mathrm{cube}\in\mathcal{M}_3}
   \left(a\cup f+2\pi z\cup a\right)\right]
   \mathcal{Z}_{\mathcal{M}_3}[z|_{\mathcal{M}_3}]
\notag\\
   &=
   \left\langle
   \Tilde{U}_{2\pi p/N}(\widetilde{M}_3)\mathcal{O}
   \right\rangle_{\mathrm{F}},
\label{eq:(C22)}
\end{align}
where we have noted that the anomalous Ward--Takahashi
identity~\eqref{eq:(C3)} applied to the present small deformation implies,
$\langle\mathcal{O}\rangle_{\mathrm{F}}^{\widetilde{M}_3'}
=\langle\mathcal{O}\rangle_{\mathrm{F}}^{\widetilde{M}_3}
\exp[ip/(8\pi N)\sum_{\mathrm{hypercube}\in\mathcal{V}_4}f\cup f]$. This
shows that the symmetry operator is actually topological.

Under the 0-form $U(1)$ lattice gauge transformation,
\begin{equation}
   u(x,\mu)\to e^{i\phi(x)}u(x,\mu)e^{-i\phi(x+\Hat{\mu})},\qquad
   -\pi<\phi(x)\leq\pi,
\label{eq:(C23)}
\end{equation}
the gauge potential in~Eq.~\eqref{eq:(2.12)} changes as
\begin{equation}
   a_\mu(x)\to a_\mu(x)-\partial_\mu\phi(x)-2\pi l_\mu(x)
   \Rightarrow a\to a-\delta\phi-2\pi l,
\label{eq:(C24)}
\end{equation}
where $l\in C^1(\Gamma,\mathbb{Z})$. Since the field strength~\eqref{eq:(2.13)}
is gauge invariant, the gauge transformation of~$z$ is given by
\begin{equation}
   z\to z+\delta l.
\label{eq:(C25)}
\end{equation}
From~Eq.~\eqref{eq:(C8)}, we find that
$\mathcal{Z}_{\mathcal{M}_3}[z|_{\mathcal{M}_3}]$ changes under the gauge
transformation as
\begin{equation}
   \mathcal{Z}_{\mathcal{M}_3}[z|_{\mathcal{M}_3}]
   \to
   \exp
   \left[
   -\frac{ip\pi}{N}\sum_{\mathrm{cube}\in\mathcal{M}_3}
   \left(
   z\cup l+l\cup z+\delta l\cup l
   \right)
   \right]
   \mathcal{Z}_{\mathcal{M}_3}[z|_{\mathcal{M}_3}].
\label{eq:(C26)}
\end{equation}
On the other hand, under the Bianchi identities, $\delta z=\delta f=0$,
\begin{align}
   &\exp
   \left[
   -\frac{ip}{4\pi N}\sum_{\mathrm{cube}\in\mathcal{M}_3}
   \left(a\cup f+2\pi z\cup a\right)\right]
\notag\\
   &\to\exp
   \left[
   -\frac{ip}{4\pi N}\sum_{\mathrm{cube}\in\mathcal{M}_3}
   \left(a\cup f+2\pi z\cup a\right)
   +\frac{ip\pi}{N}\sum_{\mathrm{cube}\in\mathcal{M}_3}
   \left(z\cup l+l\cup z+\delta l\cup l\right)\right].
\label{eq:(C27)}
\end{align}
Since $\langle\mathcal{O}\rangle_{\mathrm{F}}^{\widetilde{M}_3}$
in~Eq.~\eqref{eq:(C21)} is gauge invariant for a gauge-invariant
operator~$\mathcal{O}$ as we have demonstrated in the main text, the symmetry
operator defined by~Eq.~\eqref{eq:(C21)} is manifestly invariant under
arbitrary 3D lattice gauge transformations.

Some remarks are in order. First, the continuum BF theory corresponding
to~Eq.~\eqref{eq:(C8)} would be given by
\begin{equation}
   \mathcal{Z}_{\mathcal{M}_3}[z]
   =\int\mathrm{D}[b]\mathrm{D}[c]\,
   \exp\left\{
   \frac{ip\pi}{N}\int_{\mathcal{M}_3}
   \left[b(dc-z)-zc\right]\right\}.
\label{eq:(C28)}
\end{equation}
This system gives rise to the 't~Hooft anomaly under the background 1-form
gauge transformation, $z\to z+dl$,
\begin{equation}
   \mathcal{Z}_{\mathcal{M}_3}[z+dl]
   =\mathcal{Z}_{\mathcal{M}_3}[z]
   \exp\left[-\frac{ip\pi}{N}\int_{\mathcal{M}_3}(lz+zl+ldl)\right].
\label{eq:(C29)}
\end{equation}
This anomaly is reproduced by a 4D symmetry-protected topological (SPT) action,
\begin{equation}
   \exp\left(-\frac{ip\pi}{N}\int_{\mathcal{V}_4}zz\right),
\label{eq:(C30)}
\end{equation}
assuming that $\mathcal{M}_3=\partial(\mathcal{V}_4)$. From this anomaly
inflow, one may expect the relation~\eqref{eq:(C20)}, for which we provided a
precise proof on the lattice.

Another point is the case in which the integers $p$ and~$N$ are not co-prime
and $p=kp'$ and~$N=kN'$ with the greatest common divisor~$k$. Since the chiral
rotation angle~\eqref{eq:(C1)} depends only on the ratio,
$\alpha=2\pi p/N=2\pi p'/N'$, we may expect a certain relation between
$\mathcal{Z}_{\mathcal{M}_3}^{(p,N)}[z]$
and~$\mathcal{Z}_{\mathcal{M}_3}^{(p',N')}[z]$ for such a case. Noting that
\begin{equation}
   \int\mathrm{D}[b]
   \exp\left[\frac{ip'\pi}{N'}\sum_{\mathrm{cube}\in\mathcal{M}_3}b(\delta c-z)
   \right]
   =\delta_{N'}[\delta c-z],
\label{eq:(C31)}
\end{equation}
if we employ~Eq.~\eqref{eq:(C6)} as it stands also for this case,
Eqs.~\eqref{eq:(C8)} and~\eqref{eq:(C14)} are modified as (here
$\nu\in C^1(\mathcal{M}_3;\mathbb{Z}_{N'})$),
\begin{align}
   \mathcal{Z}_{\mathcal{M}_3}^{(p,N)}[z]
   &=\frac{1}{|C^0(\mathcal{M}_3;\mathbb{Z}_N)|}
   \int\mathrm{D}[c]\,
   \delta_{N'}\left[\delta c-z\right]
   \exp\left(
   -\frac{ip'\pi}{N'}
   \sum_{\mathrm{cube}\in\mathcal{M}_3}z\cup c\right)
\notag\\
   &=\frac{1}{|C^0(\mathcal{M}_3;\mathbb{Z}_N)|}
   \int\mathrm{D}[c]\,
   \delta_{N'}\left[\delta c-\delta\nu\right]
   \exp\left(
   -\frac{ip'\pi}{N'}
   \sum_{\mathrm{cube}\in\mathcal{M}_3}\delta\nu\cup c\right)
\notag\\
   &=\frac{1}{|C^0(\mathcal{M}_3;\mathbb{Z}_N)|}
   \int\mathrm{D}[c]\,
   \delta_{N'}\left[\delta c\right]
   \exp\left(
   -\frac{ip'\pi}{N'}
   \sum_{\mathrm{cube}\in\mathcal{M}_3}\delta\nu\cup\nu\right)
\notag\\
   &=\frac{1}{N^s}N^l\frac{1}{(N')^{f-c+1}}(N')^{b_2}
   \exp\left(
   -\frac{ip'\pi}{N'}
   \sum_{\mathrm{cube}\in\mathcal{M}_3}\delta\nu\cup\nu\right)
\notag\\
   &=\frac{1}{k^{s-l}}\mathcal{Z}_{\mathcal{M}_3}^{(p',N')}[z],
\label{eq:(C32)}
\end{align}
where we have repeated the calculation in~Eq.~\eqref{eq:(C11)}. This relation,
however, is not satisfactory in that it depends on the way of discretization
of~$\mathcal{M}_3$ through the number~$s-l$. We may evade this point by
generalizing the definition~\eqref{eq:(C6)} to
\begin{equation}
   \mathcal{Z}_{\mathcal{M}_3}[z]
   =k^{s-l}\frac{1}{|C^0(\mathcal{M}_3;\mathbb{Z}_N)|}
   \int\mathrm{D}[b]\mathrm{D}[c]\,
   e^{-S_{\mathrm{BF}}},
\label{eq:(C33)}
\end{equation}
where $k$ is the greatest common divisor of~$p$ and~$N$. Then, we obtain a
simple equality,
$\mathcal{Z}_{\mathcal{M}_3}^{(p,N)}[z]=\mathcal{Z}_{\mathcal{M}_3}^{(p',N')}[z]$.

\subsection{Fusion rules}
\label{sec:C.2}
As an application of our representation~\eqref{eq:(C21)}, we evaluate fusion
rules of two symmetry operators sharing a common 3-cycle~$\mathcal{M}_3$.
From~Eq.~\eqref{eq:(C8)}, we have (here we assume that $\gcd(p_1+p_2,N)=1$)
\begin{align}
   &\mathcal{Z}_{\mathcal{M}_3}^{(p_1,N)}[z]
   \mathcal{Z}_{\mathcal{M}_3}^{(p_2,N)}[z]
\notag\\
   &=\frac{1}{|C^0(\mathcal{M}_3;\mathbb{Z}_N)|^2}
   \int\mathrm{D}[c_1]\,
   \int\mathrm{D}[c_2]\,
   \delta_N\left[\delta c_1-z\right]
   \delta_N\left[\delta c_2-z\right]
\notag\\
   &\qquad{}
   \times\exp\left[
   -\frac{ip_1\pi}{N}
   \sum_{\mathrm{cube}\in\mathcal{M}_3}z\cup(c_1-c_2)
   -\frac{i(p_1+p_2)\pi}{N}
   \sum_{\mathrm{cube}\in\mathcal{M}_3}z\cup c_2\right]
\notag\\
   &=\mathcal{Z}_{\mathcal{M}_3}^{(p_1,N)}[0]
   \mathcal{Z}_{\mathcal{M}_3}^{(p_1+p_2,N)}[z]
\notag\\
   &=\mathcal{Z}_{\mathcal{M}_3}[0]
   \mathcal{Z}_{\mathcal{M}_3}^{(p_1+p_2,N)}[z],
\label{eq:(C34)}
\end{align}
where, in the last line, we have noted that
$\mathcal{Z}_{\mathcal{M}_3}^{(p_1,N)}[0]$ is given
by~$\mathcal{Z}_{\mathcal{M}_3}[0]$ in~Eq.~\eqref{eq:(C11)}. In the second
equality, we have noted that if
$z\neq 0\bmod N\in H^2(\mathcal{M}_3;\mathbb{Z}_N)$, the left-hand side
identically vanishes and the right-hand side does too; the equality therefore
holds. Otherwise, if $z=0\bmod N\in H^2(\mathcal{M}_3;\mathbb{Z}_N)$, there
exists $\nu\in C^1(\mathcal{M}_3;\mathbb{Z}_N)$ such that $z=\delta\nu\bmod N$
and then $\sum_{\mathrm{cube}\in\mathcal{M}_3}z\cup(c_1-c_2)
=\sum_{\mathrm{cube}\in\mathcal{M}_3}\nu\cup\delta(c_1-c_2)=0\bmod N$ under the
delta functional. In the symmetry operator~\eqref{eq:(C21)}, the factors other
than $\mathcal{Z}_{\mathcal{M}_3}[z|_{\mathcal{M}_3}]$ obey the ordinary
multiplication law. Hence, the fusion rule of the symmetry operators reads
\begin{equation}
   \Tilde{U}_{2\pi p_1/N}\Tilde{U}_{2\pi p_2/N}
   =\mathcal{Z}_{\mathcal{M}_3}[0]\Tilde{U}_{2\pi(p_1+p_2)/N}.
\label{eq:(C35)}
\end{equation}
The fact that this fusion depends on the partition function of a TQFT,
$\mathcal{Z}_{\mathcal{M}_3}[0]$~\eqref{eq:(C11)}, shows the noninvertibility.

Setting $p_2=-p_1=p$ in~Eq.~\eqref{eq:(C34)}, we have the condensation
operator~$\mathcal{C}_{\mathcal{M}_3}[z]$,
\begin{align}
   \mathcal{Z}_{\mathcal{M}_3}^{(p,N)}[z]
   \mathcal{Z}_{-\mathcal{M}_3}^{(p,N)}[z]
   &=\mathcal{Z}_{\mathcal{M}_3}^{(p,N)}[z]
   \mathcal{Z}_{\mathcal{M}_3}^{(-p,N)}[z]
\notag\\
   &=\mathcal{Z}_{\mathcal{M}_3}[0]
   \mathcal{C}_{\mathcal{M}_3}[z],
\label{eq:(C36)}
\end{align}
where
\begin{align}
   \mathcal{C}_{\mathcal{M}_3}[z]
   &=\mathcal{Z}_{\mathcal{M}_3}^{(0,N)}[z]
\notag\\
   &=\frac{1}{|C^0(\mathcal{M}_3;\mathbb{Z}_N)|}
   \int\mathrm{D}[c]\,
   \delta_N\left[\delta c-z\right]
\notag\\
   &=\begin{cases}
   \mathcal{Z}_{\mathcal{M}_3}[0]&
   \text{for $z=0\bmod N\in H^2(\mathcal{M}_3;\mathbb{Z}_N)$},\\
   0&\text{otherwise}.\\
   \end{cases}
\label{eq:(C37)}
\end{align}
The condensation operator~$\mathcal{C}_{\mathcal{M}_3}[z]$ is equal to the
topological factor~$\mathcal{Z}_{\mathcal{M}_3}[0]$ with the projection operator
which restricts the $z$~flux to multiples of~$N$.

A more general fusion can be obtained by considering (here we assume
that $\gcd(p_1,N_1)=\gcd(p_2,N_2)=1$)
\begin{align}
   &\mathcal{Z}_{\mathcal{M}_3}^{(p_1,N_1)}[z]
   \mathcal{Z}_{\mathcal{M}_3}^{(p_2,N_2)}[z]
\notag\\
   &=\frac{1}{|C^0(\mathcal{M}_3;\mathbb{Z}_{N_1})|}
   \frac{1}{|C^0(\mathcal{M}_3;\mathbb{Z}_{N_2})|}
   \int\mathrm{D}[c_1]\,
   \delta_{N_1}\left[\delta c_1-z\right]
   \int\mathrm{D}[c_2]\,
   \delta_{N_2}\left[\delta c_2-z\right]
\notag\\
   &\qquad{}
   \times\exp\left(
   -\frac{ip_1\pi}{N_1}
   \sum_{\mathrm{cube}\in\mathcal{M}_3}z\cup c_1
   -\frac{ip_2\pi}{N_2}
   \sum_{\mathrm{cube}\in\mathcal{M}_3}z\cup c_2\right).
\label{eq:(C38)}
\end{align}
This identically vanishes unless $z=0\bmod N_1$
in~$Z^2(\mathcal{M}_3;\mathbb{Z}_{N_1})$
and $z=0\bmod N_2$ in~$Z^2(\mathcal{M}_3;\mathbb{Z}_{N_2})$. Thus, let us assume
these are matched; this implies that $z=0\bmod N$
in~$Z^2(\mathcal{M}_3;\mathbb{Z}_N)$, where $N=\lcm(N_1,N_2)$. Then, there
exists $\nu\in C^1(\mathcal{M}_3;\mathbb{Z}_N)$ such that $z=\delta\nu\bmod N$.
Using this in~Eq.~\eqref{eq:(C38)}, after the shift of variables,
$c_1\to c_1+\nu$ and~$c_2\to c_2+\nu$, we have
\begin{align}
   &\mathcal{Z}_{\mathcal{M}_3}^{(p_1,N_1)}[z]
   \mathcal{Z}_{\mathcal{M}_3}^{(p_2,N_2)}[z]
\notag\\
   &=\frac{1}{|C^0(\mathcal{M}_3;\mathbb{Z}_{N_1})|}
   \frac{1}{|C^0(\mathcal{M}_3;\mathbb{Z}_{N_2})|}
   \int\mathrm{D}[c_1]\,
   \delta_{N_1}\left[\delta c_1\right]
   \int\mathrm{D}[c_2]\,
   \delta_{N_2}\left[\delta c_2\right]
\notag\\
   &\qquad{}
   \times\exp\left[
   -\left(\frac{ip_1\pi}{N_1}+\frac{ip_2\pi}{N_2}\right)
   \sum_{\mathrm{cube}\in\mathcal{M}_3}\delta\nu\cup\nu
   \right].
\label{eq:(C39)}
\end{align}
On the other hand, setting $\beta_1=N/N_1$ and~$\beta_2=N/N_2$
for~$N=\lcm(N_1,N_2)$, and assuming that $\gcd(\beta_1p_1+\beta_2p_2,N)=1$,
\begin{align}
   &\mathcal{Z}_{\mathcal{M}_3}^{(\beta_1p_1+\beta_2p_2,N)}[z]
\notag\\
   &=\frac{1}{|C^0(\mathcal{M}_3;\mathbb{Z}_{N})|}
   \int\mathrm{D}[c]\,
   \delta_N\left[\delta c-z\right]
   \exp\left[
   -\left(\frac{ip_1\pi}{N_1}+\frac{ip_2\pi}{N_2}\right)
   \sum_{\mathrm{cube}\in\mathcal{M}_3}z\cup c
   \right].
\label{eq:(C40)}
\end{align}
This identically vanishes unless $z=0\bmod N$
in~$Z^2(\mathcal{M}_3;\mathbb{Z}_N)$. Assuming the latter, we have
\begin{align}
   &\mathcal{Z}_{\mathcal{M}_3}^{(\beta_1p_1+\beta_2p_2,N)}[z]
\notag\\
   &=\frac{1}{|C^0(\mathcal{M}_3;\mathbb{Z}_{N})|}
   \int\mathrm{D}[c]\,
   \delta_N\left[\delta c\right]
   \exp\left[
   -\left(\frac{ip_1\pi}{N_1}+\frac{ip_2\pi}{N_2}\right)
   \sum_{\mathrm{cube}\in\mathcal{M}_3}\delta\nu\cup\nu
   \right].
\label{eq:(C41)}
\end{align}
From Eqs.~\eqref{eq:(C39)} and~\eqref{eq:(C41)}, we have the fusion,
\begin{equation}
   \mathcal{Z}_{\mathcal{M}_3}^{(p_1,N_1)}[z]
   \mathcal{Z}_{\mathcal{M}_3}^{(p_2,N_2)}[z]
   =\frac{\mathcal{Z}_{\mathcal{M}_3}^{(0,N_1)}[0]
   \mathcal{Z}_{\mathcal{M}_3}^{(0,N_2)}[0]}
   {\mathcal{Z}_{\mathcal{M}_3}^{(0,N)}[0]}
   \mathcal{Z}_{\mathcal{M}_3}^{(\beta_1p_1+\beta_2p_2,N)}[z],
\label{eq:(C42)}
\end{equation}
and this shows
\begin{align}
   \Tilde{U}_{2\pi p_1/N_1}\Tilde{U}_{2\pi p_2/N_2}
   &=\frac{\mathcal{Z}_{\mathcal{M}_3}^{(0,N_1)}[0]
   \mathcal{Z}_{\mathcal{M}_3}^{(0,N_2)}[0]}
   {\mathcal{Z}_{\mathcal{M}_3}^{(0,N)}[0]}
   \Tilde{U}_{2\pi(p_1/N_1+p_2/N_2)}
\notag\\
   &=\left(\frac{N_1N_2}{N}\right)^{b_2-1}
   \Tilde{U}_{2\pi(p_1/N_1+p_2/N_2)}.
\label{eq:(C43)}
\end{align}




%



\let\doi\relax









\end{document}